\documentclass[10pt]{iopart}

\usepackage{iopams}
\expandafter\let\csname equation*\endcsname\relax
\expandafter\let\csname endequation*\endcsname\relax
\usepackage{amsmath}
\usepackage{amssymb}
\usepackage{bm}
\usepackage{tabularx}

\renewcommand{\vec}[1]{\bm{#1}}
\renewcommand{\Re}{\mathop{\mathrm{Re}}}
\renewcommand{\Im}{\mathop{\mathrm{Im}}}
\newcommand{\sgn}{\mathop{\mathrm{sgn}}}
\newcommand{\dd}{d}

\usepackage[pdfborder={0 0 0}]{hyperref}
\usepackage{graphicx}
\usepackage{grffile}

\graphicspath{{./figs/}}

\usepackage[usenames,dvipsnames]{xcolor}

\begin{document}

\title[SF hybrids for non-reciprocal electronics and detectors]{Superconductor-ferromagnet hybrids for
non-reciprocal electronics and detectors}

\newcommand{\orcid}[1]{\href{https://orcid.org/#1}{\raisebox{2.5pt}{\includegraphics[width=7pt]{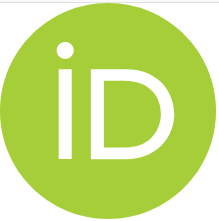}}}}

\author{
   Zhuoran Geng$^1$,  Alberto Hijano$^{2,3}$\orcid{0000-0002-3018-4395}, Stefan Ili\'{c}$^2$, Maxim Ilyn$^2$,   Ilari J. Maasilta$^1$, Alessandro Monfardini$^4$,  Maria Spies$^5$\orcid{0000-0002-3570-3422}, Elia Strambini$^5$\orcid{0000-0003-1135-2004},
  Pauli Virtanen$^1$\orcid{0000-0001-9957-1257},  Martino Calvo$^4$, Carmen Gonz\'alez-Orell\'ana$^2$,
  Ari P. Helenius$^1$, Sara Khorshidian$^5$\orcid{0000-0002-2374-0728}, Clodoaldo I. L. de Araujo$^{5,6}$\orcid{0000-0003-2801-1759}, Florence Levy-Bertrand$^4$\orcid{0000-0003-4244-1368}, Celia Rogero$^{2,7}$\orcid{0000-0002-2812-8853}, F. Giazotto$^5$\orcid{0000-0002-1571-137X}, F. Sebastián Bergeret$^{2,7}$,
  and
  Tero T. Heikkilä$^1$\orcid{0000-0002-7732-691X}
}
\address{$^1$ Department of Physics and Nanoscience Center, University of Jyv\"askyl\"a, P.O. Box 35 (YFL), FI-40014 University of Jyv\"askyl\"a, Finland}
\address{$^2$ Centro de F\'isica de Materiales (CFM-MPC) Centro Mixto CSIC-UPV/EHU, E-20018 Donostia-San Sebasti\'an,  Spain}
\address{$^3$ Department of Condensed Matter Physics, University of the Basque Country UPV/EHU, 48080 Bilbao, Spain}
\address{$^4$ Univ. Grenoble Alpes, CNRS, Grenoble INP, Institut Néel, 38000 Grenoble, France}
\address{$^5$ NEST, Istituto Nanoscienze-CNR and Scuola Normale Superiore, I-56127 Pisa, Italy}
\address{$^6$ Laboratory of Spintronics and Nanomagnetism (LabSpiN), Departamento de F\'{i}sica,
  Universidade Federal de Vi\c cosa, Vi\c cosa, 36570-900, Minas Gerais, Brazil }
\address{$^7$ Donostia International Physics Center (DIPC), 20018 Donostia-San Sebastian, Spain}

\ead{zhuoran.z.geng@jyu.fi}
\ead{maxim.ilin@ehu.es}
\ead{alessandro.monfardini@neel.cnrs.fr}
\ead{elia.strambini@cnr.it}
\ead{pauli.t.virtanen@jyu.fi}
\ead{fs.bergeret@csic.es}
\ead{tero.t.heikkila@jyu.fi}

\ioptwocol

\begin{abstract}
We review the use of hybrid thin films composed of superconductors and ferromagnets for creating non-reciprocal electronic components and self-biased detectors of electromagnetic radiation. We begin by introducing the theory behind these effects, as well as discussing various potential materials that can be used in the fabrication of these components. We then proceed with a detailed discussion on the fabrication and characterization of Al/EuS/Cu and EuS/Al/Co-based detectors, along with their noise analysis. Additionally, we suggest some approaches for multiplexing such self-biased detectors. 
\end{abstract}

\tableofcontents

\section{Introduction}

Conventional superconducting electronics \cite{braginski2019superconductor} rely on a combination of supercurrent and quasiparticle current transport across superconducting wires and different types of weak links. These combinations enable various functionalities, such as magnetometry \cite{kirtley1995high}, current or voltage amplifiers \cite{macklin2015near}, voltage standards \cite{hamilton2000josephson}, and detectors based on resistance \cite{Irwin2005} or inductance \cite{sergeev2002ultrasensitive} that depend on the nonequilibrium state of the system. Compared to their semiconductor counterparts, superconducting electronics lack a basic element: non-reciprocal devices such as diodes or thermoelectric elements. The absence of non-reciprocity can be attributed to the intrinsic electron-hole symmetry of the superconducting state. However, this symmetry can be broken using combinations of magnetic and superconducting elements \cite{ozaeta_predicted_2014,bergeret2018colloquium}, which allow, in principle, the achievement of strong non-reciprocity or thermoelectric figure of merit. These phenomena can be employed to create superconducting spintronic tunnel diodes \cite{strambini2022superconducting}, building blocks for   superconducting  logic and cryogenic memory,   or novel types of detectors, such as the superconductor--ferromagnet thermoelectric detector (SFTED)\cite{Heikkila2018}, with applications  in astrophysics for the detection of the cosmic microwave background \cite{hanson2013detection}, and terahertz-radiation sensing used, for example, in security imaging \cite{luukanen2012millimeter}.  Remarkably, in the SFTED  the absorbed radiation directly generates the desired measurement signal, without the need for a separate bias current or voltage.

This review is divided into two parts. The first part focuses on the fundamentals of the building blocks for the SFTED, namely superconductor-ferromagnetic-insulator (S/FI) bilayers. We begin with a brief account of the underlying theory, focusing on aspects of the physics that are important for functionalities.  The full  theory for the underlying transport phenomena has been summarized elsewhere \cite{bergeret2018colloquium,heikkila2019thermal}. We then  discuss the main material combinations used so far, along with their basic properties, and the characterization of the superconducting state in the presence of the magnetic proximity effect, as well as the basic non-reciprocal current-voltage characteristics and thermoelectric signals. The second part focuses on the realization of radiation sensors operating in two different ranges of detected electromagnetic signals, and  their readout.  We provide an outlook and present some open challenges in the last section.

\section{Fundamentals}

In this first part of the article, we concentrate on the fundamental aspects related to the fabrication of the SFTED. It comprises three sections that present the theory underlying thermoelectricity and non-reciprocal transport in superconductor-ferromagnetic insulator systems, the fabrication process of these building blocks, and their spectral characterization.

\subsection{Non-reciprocal transport in hybrid superconductor/ferromagnet structures}\label{sec1}

\label{subs:tunneling}

The aim of this work is to identify and realize two types of functionalities in an electronic device at low temperatures. The first of them is \textit{non-reciprocity}. Non-reciprocal electronic transport in solid-state devices implies an asymmetry in the current-voltage characteristic: $I(V)\neq -I(-V)$. Non-reciprocity is associated with inversion symmetry breaking, meaning that non-reciprocal elements usually consist of hybrid structures involving different materials.

\begin{figure}[t]
  \centering
  \includegraphics[width=\columnwidth]{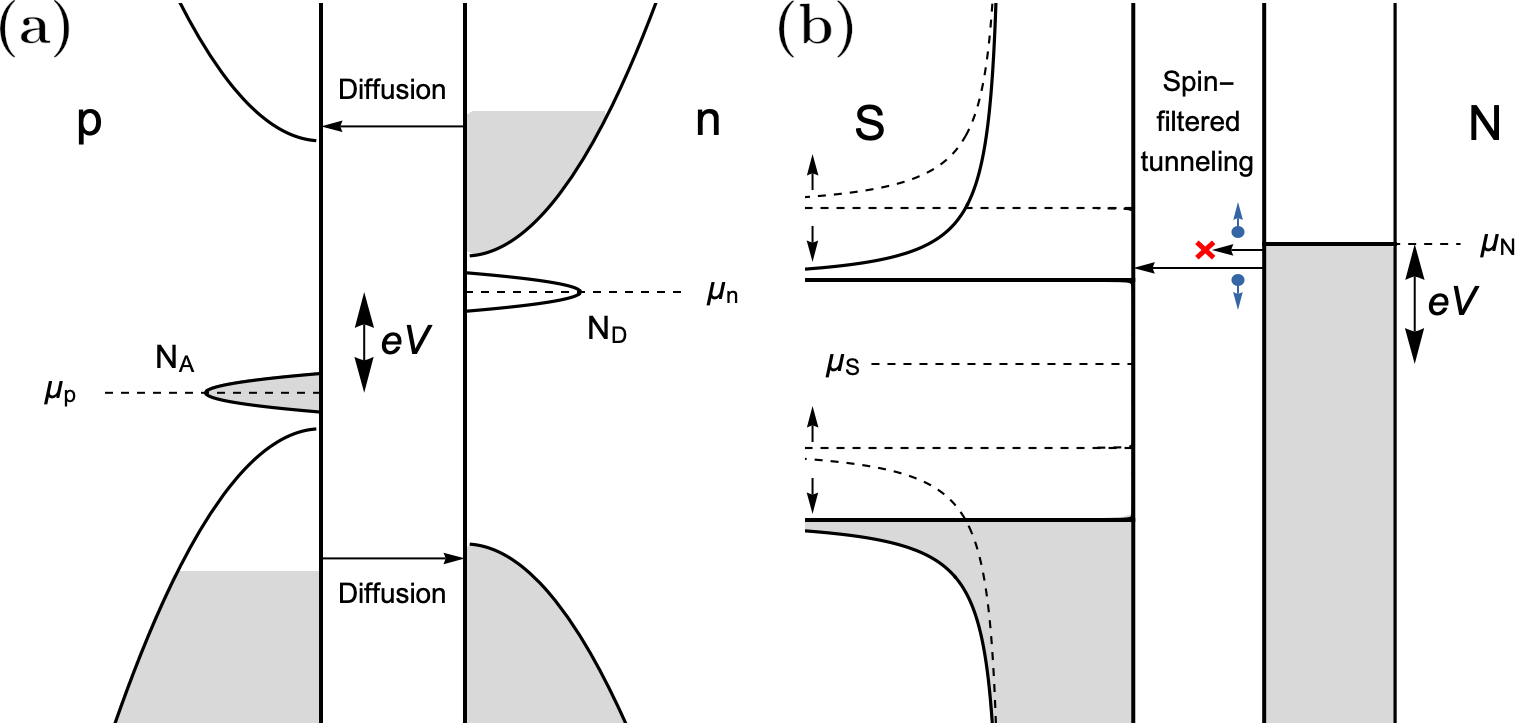}
  \caption{
    \label{fig:tunneling}
    Density of states (solid line) and charge carrier density (shaded region) in thermal equilibrium for a (a) p-n junction and (b) spin-split S/N tunnel junction. The dashed lines in panel (b) represent the density of states of the filtered states. Here $\mu_p$ and $\mu_n$ are the quasi-chemical potentials of the p- and n-regions, and $\mathrm{N_A}$ and $\mathrm{N_D}$ are the holes and electrons supplied by the acceptor and donor impurities, respectively.   }
\end{figure}

A typical example is provided by p-n diodes, which consist of p-type and n-type semiconducting layers in contact with each other. In addition to the broken inversion symmetry, the non-reciprocal electron currents require the breaking of the electron-hole symmetry resulting from the n- and p-doping of the two semiconducting layers. The working principle of a p-n diode can be understood from the sketch in Fig.~\ref{fig:tunneling}(a), which shows a positively biased p-n junction.
In the p-n junction, free electrons from the n-region diffuse into the p-region and vice versa for the holes. The regions near the interface lose their charge neutrality, with the negatively charged acceptor dopant atoms remaining in the p-region and the positively charged donor dopants remaining in the n-region, forming the depletion layer. This charged region establishes a built-in potential that counteracts quasiparticle diffusion, leading to zero net current at equilibrium. A forward bias allows the majority carriers of each region to cross the depletion layer and be injected into the nearby region, where they recombine with opposite charge quasiparticles.

For several electronics-based quantum technologies, the interesting operating regime lies at sub-Kelvin temperatures. This poses certain challenges to traditional semiconductor technology, and substantial effort has been dedicated to scaling this technology down to the lowest temperatures \cite{ghibaudo2001,gutierrez2001,gonzalez2021}. Due to these challenges, research into new alternatives for non-reciprocal electronic transport at low temperatures has intensified.

An alternative realization of non-reciprocity to traditional semiconducting devices can be achieved by combining two types of correlated interacting electronic phases: superconductivity and magnetism. This review concentrates on hybrid heterostructures formed using these elements. We illustrate the working principle of a superconducting spintronic tunnel diode in Fig.~\ref{fig:tunneling}(b). There, the spin-split superconducting state along with spin-polarized tunneling conspire to provide unidirectional current. The quasiparticles pass through the insulating barrier due to quantum tunneling. In the absence of spin splitting, at $T=0$, there is no tunneling current until the applied bias is greater than the superconducting gap $e|V|>\Delta$, since the chemical potential difference must provide enough energy to create an excitation in the superconductor. If particle-hole symmetry is broken due to spin-splitting and spin filtering, for instance, the bias required for a finite current in the forward bias configuration $eV>\Delta-h$ is smaller than in the backward bias $eV<-\Delta-h$ ($h$ is the exchange field), so that the charge transport is non-reciprocal.

The second functionality we discuss is \textit{thermoelectricity}, where linear electric excitation leads to heat currents, or temperature differences can be associated with charge currents. In other words, the linear response charge and heat currents $I,\dot Q$ can be expressed in terms of bias voltage $V$ and temperature difference $\Delta T$ as
\begin{equation}
\begin{pmatrix} I \\ \dot Q \end{pmatrix}  = \begin{pmatrix} G & \alpha \\ \alpha' & G_{\rm th} T \end{pmatrix} \begin{pmatrix} V \\ \Delta T/T\end{pmatrix},
\end{equation}
where $G$ and $G_{\rm th}$ are charge and heat conductances, respectively, and $\alpha$ is the thermoelectric coefficient. By Onsager symmetry, $\alpha'$ is obtained from $\alpha$ through the time-reversal transformation. In general, $\alpha$ characterizes the degree of electron-hole symmetry breaking in the spectrum. When $\alpha'=\alpha$ (appropriate for this paper), the thermoelectric effects are characterized by the dimensionless figure of merit $ZT=\alpha^2/(G_{\rm th} G  T-\alpha^2)$. In ordinary metals with a large Fermi energy ($E_F$) it is of the order of $o(k_B T/E_F)$, and therefore very small at sub-Kelvin temperatures. Semiconductor materials exhibit the most significant thermoelectric effects at around room temperature \cite{zhao2014ultralow}. However, due to their 10-100 meV scale energy gap, semiconductors tend to freeze out at sub-Kelvin temperatures, and the thermoelectric efficiency is lost. Similar to the diode functionality, sub-Kelvin thermoelectricity can be realized through the combination of superconductors and ferromagnets, as detailed in the following sections.

\begin{figure}[t]
\centering
\includegraphics[width=0.55\columnwidth]{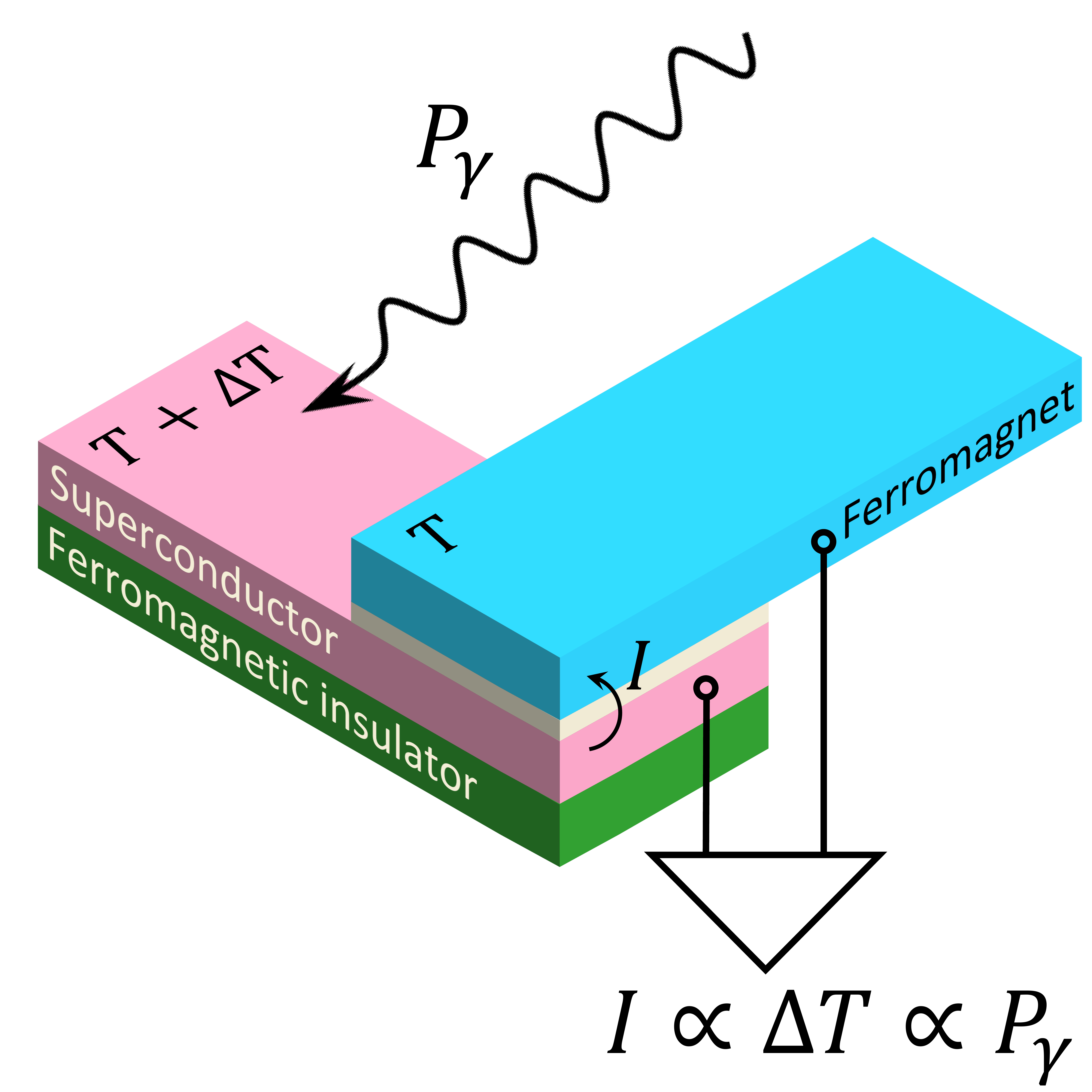}
\caption{Schematic idea for biasless detection based on thermoelectricity. Absorbed light with power $P_\gamma$ heats up one part of the heterostructure, creating a temperature difference $\Delta T$, which results in a thermoelectric current $I=\alpha \Delta T/T \propto P_\gamma$.}
\label{fig:biaslessdetection}
\end{figure}

Both non-reciprocity and thermoelectricity can be utilized in detectors. The majority of existing superconducting detector types, such as transition edge sensor (TES) \cite{Irwin1996,Irwin2005} and kinetic inductance detector (KID) \cite{Grossman1991} are based on probing how the response of the material changes upon irradiation. This requires the use of additional probing lines, which can also introduce added noise. Moreover, such detectors tend to be rather non-linear and therefore they can saturate relatively quickly. Conversely, using a diode or a thermoelectric element as a sensor enables converting the detected radiation directly into a current or voltage. Hence, such detectors are \textit{self-biased} by the radiation (as in Fig.~\ref{fig:biaslessdetection}).

In the following, we describe how the spin-dependent scattering at an interface between a superconductor (S) and a ferromagnetic insulator (FI) leads to two key phenomena underlying both the non-reciprocity and thermoelectric transport in such systems: spin splitting of the density of states and spin-polarized transport.

\subsubsection{S/FI layer structures}
\label{subsec:SFI}

The magnetic proximity effect concept in hybrid systems, comprising ferromagnetic insulators and superconductors, was introduced early on by de Gennes \cite{degennes1966-cfs} within the realm of superconductivity theory.
The first experimental observations of spin splitting associated with the interfacial exchange interaction in superconductors were, however, obtained much later in  Refs.~\cite{tedrow_spin-polarized_1986,tkaczyk_magnetic_1987} using EuO/Al systems, followed by observations in EuS/Al systems, also at zero external field \cite{moodera_electron-spin_1988}. Subsequently, other material combinations have been explored, and the theoretical understanding has been refined.

\begin{figure}[t]
  \centering
  \includegraphics[width=0.75\columnwidth]{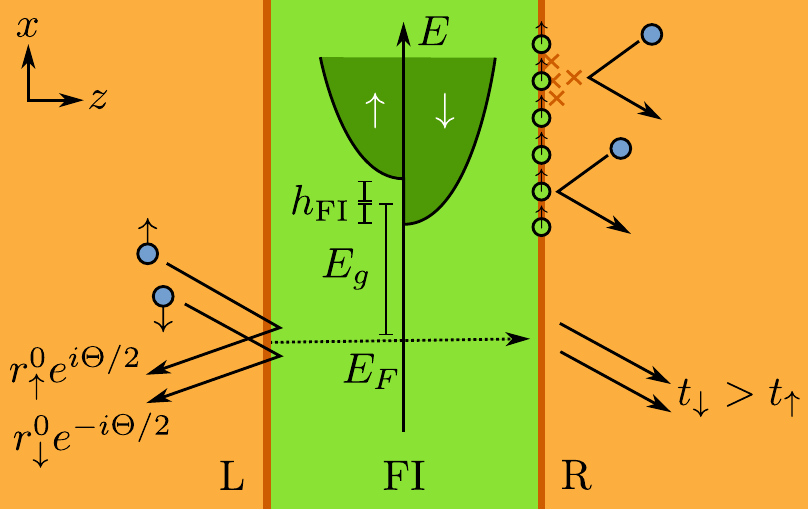}
  \caption{
    \label{fig:scattering}
    Spin-dependent scattering and tunnelling through a ferromagnetic insulator (FI) sandwiched
    between two conductors (L and R). In a simplified band model, \cite{tokuyasu_proximity_1988} the electron bands of the
    FI exhibit a gap $E_g$, and are spin split by an intrinsic exchange field $h_{\mathrm FI}$.
    Electron scattering from $L$ to $R$ at Fermi level $E_F$ is spin-dependent:
    the reflection coefficients $r$
    acquire a spin-dependent phase shift $\Theta$, and the transmission is polarized to favor
    one spin direction. Top right: 
    The magnetic proximity effect can also be interpreted through interaction with local
    moments at the FI interface.
    The physics is greatly influence by the interface quality.
  }
\end{figure}

Electron transport at ferromagnetic insulator interfaces is explained by a scattering model as introduced in Ref.~\cite{tokuyasu_proximity_1988} and further discussed in \cite{millis1988,khusainov2002}. Figure \ref{fig:scattering} illustrates this with a ferromagnetic insulator layer between two conductors. At the FI interface, an incident electron with spin $\sigma=\uparrow,\downarrow$ can reflect back, exhibiting a spin-sensitive amplitude given by $r_{\uparrow/\downarrow}= r_{\uparrow/\downarrow}^0 e^{\pm i \Theta/2}$. In a thicker FI layer, while the reflection
probabilities for both spins are nearly one, the amplitudes differ by a
scattering phase shift $e^{i\Theta}$, where $\Theta$ is called the spin-mixing angle.
If conductor L is a clean thin film of thickness $d$, electron eigenstates in it
are given by a quantization condition $e^{2ik_z(\epsilon_{\uparrow/\downarrow})d \pm i\Theta/2 + i\phi_0}=1$,
where $\phi_0$ is a spin-independent phase shift. Consequently, $\Theta$ induces a spin splitting 
\begin{align}
  \label{eq:spin-splitting-ballistic}
  \bar{h}
  &=
  \frac{\epsilon_\uparrow - \epsilon_\downarrow}{2}
  \approx
  \frac{\Theta}{4d} \left|\frac{dk_z}{d\epsilon}\right|^{-1}
  =
  \frac{v_F \cos(\alpha) \Theta}{4d}
  \,,
\end{align}
where $k_z=k_F\cos\alpha=\sqrt{2m^*\epsilon - k_\parallel^2}$ is the incident momentum of the electron hitting the FI
interface at angle $\alpha$, $|\Theta|\ll1$, and $m^*$ is the effective mass.
The splitting depends inversely on the thickness $d$\{, as predicted by de Gennes \cite{degennes1966-cfs}. A scattering calculation in such a band model, \cite{tokuyasu_proximity_1988} assuming the FI has a band gap $E_g$ and an internal exchange field $h_{\mathrm{FI}}$, 
gives the spin-mixing angle 
$\Theta\approx2[\sqrt{E_F/E_g}+\sqrt{E_g/E_F}]^{-1}(h_{\mathrm{FI}}/E_g)\cos\alpha=\Theta_0\cos\alpha$ 
for $E_g\gtrsim{}E_F$, which is proportional to $h_{\rm FI}$.
On the other hand, the transmission amplitudes
$t_{\uparrow/\downarrow}\propto{}\exp\bigl[-d_{\mathrm{FI}}\sqrt{2m^*(E_g\pm{}h_{\mathrm{FI}})+k_\parallel^2}\bigr]$
are spin-dependent and the spin-filtering polarization $\mathcal{P}=(|t_\uparrow|^2-|t_\downarrow|^2)/(|t_\uparrow|^2+|t_\downarrow|^2)$ increases by increasing the FI thickness $d_{\mathrm{FI}}$, which is also observed \cite{esaki1967,mueller1972,meservey_spin-polarized_1994,miao_tunneling_2011}.
Qualitatively, this simplified model predicts then both the presence of the magnetic proximity effect and the spin-polarized tunneling, even though its quantitative accuracy is limited \cite{tokuyasu_proximity_1988,hao1991}.

The effect of FI interfaces on superconductors can also be described by modeling the
exchange interaction between the conduction electrons and the localized magnetic moments $\vec{S}_{\vec{r}}$ at the FI interface,
\cite{khusainov1996indirect,khusainov2002}
\begin{align}\label{H_ex}
H_{\rm ex} = - J_{\rm ex} \sum_{\vec{r}\alpha\beta} \Psi^\dagger_\alpha(\vec{r})(\vec{S}_{\vec{r}}\cdot\vec{\sigma})_{\alpha\beta} \Psi_\beta(\vec{r})
\,,
\end{align}
where $J_{\rm ex}$ is an effective parameter describing the interaction strength.
Semiclassical theory suggests the generation of a localized interfacial Zeeman field for conduction electrons, represented as $h_{\rm int}(z)=h_{\rm ex}'\delta(z)$, complemented by spin-flip scattering. \cite{khusainov2002} 
Such $\delta$-function model can match experiments reasonably well \cite{strambini_revealing_2017}.
Here $h_{\mathrm{ex}}'=-J_{ex}n_m\langle{S}\rangle/\hbar$, with $n_m$ the 2D surface density of magnetic moments
and $\langle{S}\rangle$ their average spin. \cite{zhang2019theory}
In certain limits, this model can be mapped to the scattering model, as the scattering matrices
can be incorporated to boundary conditions of semiclassical transport equations \cite{millis1988,tokuyasu_proximity_1988,huertas-hernando_absolute_2002,brataas2000,eschrig2015}.
These models allow for a convenient description of FI/superconductor(S) hybrid systems, also in cases
where the superconductors are not ideal, e.g., containing potential or magnetic impurities.

One way to probe the magnetic interface properties experimentally is via spin transport measurements
\cite{tserkovnyak2005,sinova2015spin}.
In the scattering model, these properties are often characterized by the complex-valued spin-mixing conductance $G_{\uparrow\downarrow}$.
The spin-dependent current incoming to the interface is $I_{\mathrm{in}}^{\sigma\sigma'}\propto(a_{\mathrm{in}}^{\sigma'})^* a_{\mathrm{in}}^\sigma$, where
$a_{\mathrm{in}}$ are the incoming spin-dependent scattering amplitudes. The difference compared to the reflected current is \cite{brataas2000,tserkovnyak2005}
\begin{align}
  \label{eq:Isdiff}
  I_{\mathrm{in}}^{\sigma\sigma'} - I_{\mathrm{refl}}^{\sigma\sigma'}
  =
  [1 - (r_\sigma)^* r_{\sigma'}] I_{\mathrm{in}}^{\sigma\sigma'}
  =
  g_{\sigma\sigma'} I_{\mathrm{in}}^{\sigma\sigma'}
  \,.
\end{align}
The spin-mixing conductance $G_{\uparrow\downarrow}=\frac{e^2}{2\pi\hbar}\sum_{nn'}[\delta_{nn'}-(r^{nn'}_\uparrow)^*r^{nn'}_\downarrow]$ is defined by summing
$g_{\uparrow\downarrow}$ over all scattering channels $n$, and describes the 
absorbed spin current transverse to the magnetization of the FI. In the simple
interface scattering model, and assuming complete reflection $|r_\sigma|=1$, we
have $g_{\sigma\sigma}=0$ and
$g_{\uparrow\downarrow}=g^r_{\uparrow\downarrow}+ig^i_{\uparrow\downarrow}$,
$g^r_{\uparrow\downarrow}=1-\cos\Theta$,
$g^i_{\uparrow\downarrow}=\sin\Theta$. For $|\Theta|\ll1$,
$g_{\uparrow\downarrow}\approx{}i\Theta$ is imaginary,
and the spin-mixing conductance per square area is found to be
\cite{huertas-hernando_absolute_2002,eschrig2015,zhang2019theory}
\begin{align}
  \Im G_{\uparrow\downarrow}/\mathcal{A}
  \approx
  \frac{e^2}{\hbar}
  N_F h_{\rm ex}'
  \approx
  \frac{e^2}{\hbar}\frac{k_F^2}{16\pi^2}\Theta_0
  \,,
  \label{eq:Gupdownmod}
\end{align}
where $N_F$ is the Fermi level density of states per spin. That the imaginary part of $G_{\uparrow\downarrow}$
is associated with the interfacial exchange holds also more generally
\cite{huertas-hernando_absolute_2002,zhang2019theory}.
Physically, the absorbed spin current produces spin torque acting on the FI magnetization, leading to phenomena
observable for example in ferromagnetic resonance experiments; see
Ref.~\cite{tserkovnyak2005} for a review.
Furthermore, in conjunction with the spin-Hall effect, it provides an observable contribution to the magnetoresistance.\cite{nakayama2013,chen2013,sinova2015spin}

The magnetic proximity effect in S/FI bilayers can also be observed \cite{tedrow_spin-polarized_1986,hao1991,meservey_spin-polarized_1994,strambini_revealing_2017} by probing the density of states (DOS) $N_\sigma(E)$ in S, as the effective exchange field $\bar{h}$ 
from \eqref{eq:spin-splitting-ballistic} splits the BCS peaks in the DOS. In ideal superconductors,
$N_{\sigma}=N_{\mathrm{BCS},\sigma}$, the splitting is
\begin{align}
   \label{eq:DOS-splitting}
   N_{\mathrm{BCS},\sigma=\uparrow/\downarrow}(E)
   =
   \Re
   \frac{-i(E \pm \bar{h} + i0^+)}{\sqrt{\Delta^2 - (E\pm \bar{h} + i0^+)^2}}
   \,.
\end{align}
Although the splitting $\bar{h}$ in the ballistic model depends on the momentum direction,
it is expected to be averaged by disorder of the bulk and the interface, and by Fermi liquid interactions
\cite{tokuyasu_proximity_1988,hao1991,khusainov2002,virtanen2020}. For thin films, in this case one expects \cite{khusainov2002}
\begin{align}
   \label{eq:h-vs-d-diffusive}
   \bar{h} \approx \frac{h_{\rm ex}'}{d}
\end{align}
which scales similarly with film thickness $d$ as in the ballistic limit.
Experimental evidence is compatible with
momentum independent $\bar{h}$ \cite{hao1991}.

Experimentally, $N_\sigma$ can be probed via spin-polarized tunneling
\cite{meservey_spin-polarized_1994,miao_tunneling_2011}.
Tunneling current through a spin-polarized contact to a spin-split
superconductor is
\begin{align}
  I(V)
  &=
  \sum_{\sigma=\uparrow,\downarrow}
  \frac{G_\sigma}{e}
  \int_{-\infty}^\infty dE\,
  N_\sigma(E)[f_S(E) - f_N(E + eV)]
  \,,
  \label{eq:I(V)}
\end{align}
where $f_{N,S}$ are Fermi distribution functions on the superconductor/normal sides 
of the tunnel junction, and $G_\sigma$ are the spin-dependent tunneling conductances.
In S/FI/N junctions, where the ferromagnetic insulator also serves as the tunnel barrier,
the spin-dependent conductances $G_\sigma\propto{}|t_\sigma|^2$ inherit spin-dependence from the tunneling
amplitudes $t_{\uparrow/\downarrow}$. In F/I/S/FI structures, where the
tunnel electrode itself is ferromagnetic, the spin-dependence in $G_\sigma\propto{}N_{F,\sigma}$ is influenced by its spin-dependent density of states. However, interface properties generally play a significant role as well \cite{deteresa1999,miao_tunneling_2011}.

Usually, experimental $I-V$ characteristics do not match Eq.~\eqref{eq:I(V)} with the ideal BCS density of states 
of Eq.~\eqref{eq:DOS-splitting}. These discrepancies can be addressed by considering the non-idealities of the superconductor that alter $N_\sigma$, as discussed below. Other contributing factors may include the structure of the tunnel junction, such as pinholes and other irregularities, as well as features of the experimental setup like voltage fluctuations~\cite{pekola_environment-assisted_2010}.
To account for these features phenomenologically, one can introduce the Dynes parameter $\Gamma>0$ and replace $E$ with $E+i\Gamma$ in Eq.~\eqref{eq:DOS-splitting}.
We return to these points in Sec.~\ref{sec:tunnel-characteristics}.

The phase diagram of thin spin-split superconducting films has been extensively studied, {
theoretically \cite{chandrasekhar1962,clogston1962,bruno1973-mfs,khusainov2002}
and experimentally \cite{catelani_fermi-liquid_2008,xiong_spin-resolved_2011,khusainov2002}.}
A main feature at low temperatures is the presence of a first-order transition between
superconducting and normal states, which in the ideal case occurs at the Chandrasekhar--Clogston value of the exchange field
$h_c=\Delta/\sqrt{2}$. This is expected to be affected by factors such as Fermi liquid interactions, and short spin-flip $\tau_{sf}$ or spin-orbit $\tau_{so}$ scattering times in the superconductor {which} generally reduce the effect of the exchange field \cite{bruno1973-mfs,alexander1985,heikkila2019thermal} and also change the tunneling DOS.

\begin{figure}[t]
  \centering
  \includegraphics[width=0.85 \columnwidth]{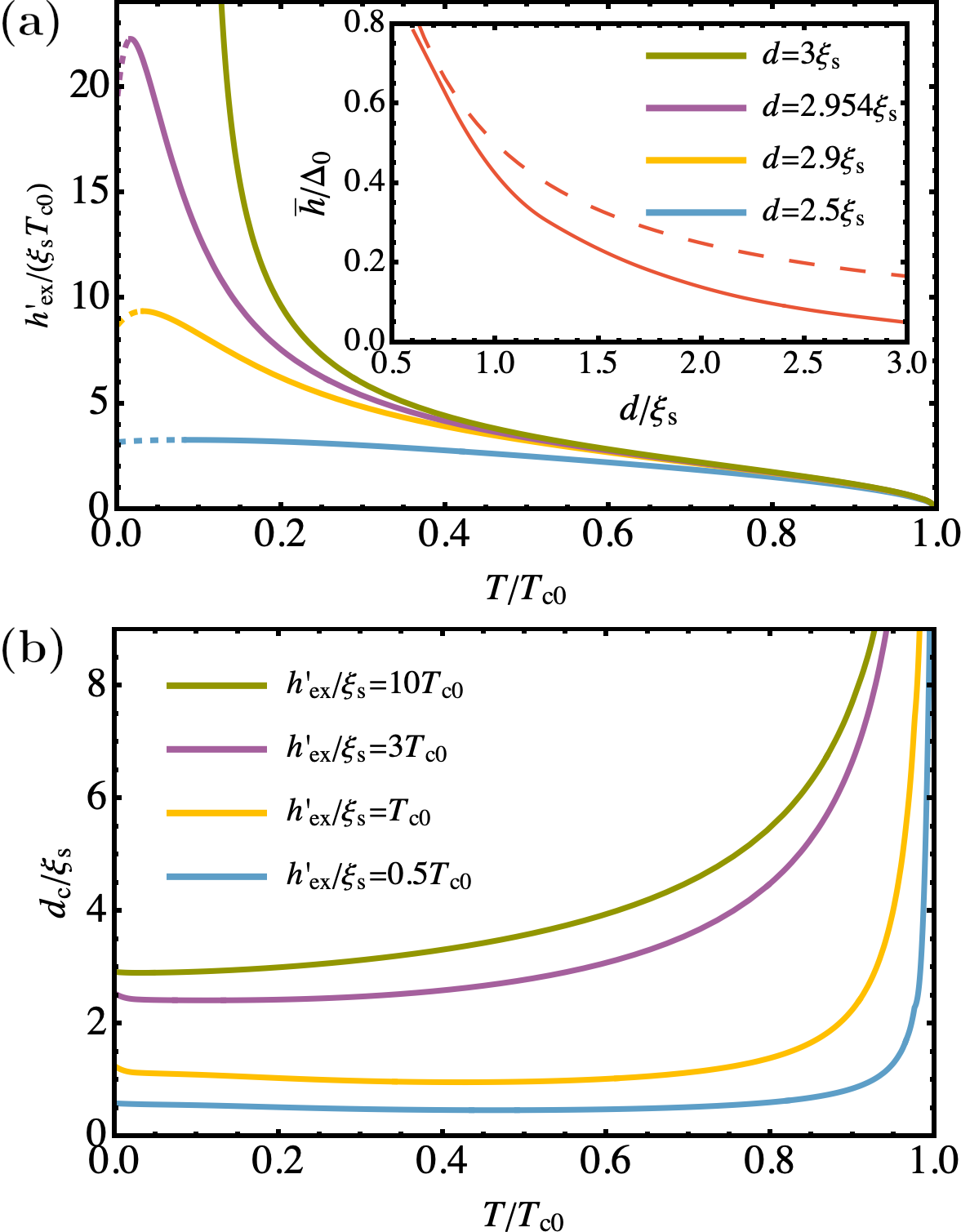}
  \caption{
    \label{fig:phase_diagram}
    (a) Critical exchange field  for different thicknesses $d$ of the superconductor  and (b) critical thickness for different interfacial exchange fields $h_{\rm ex}'$. The dashed lines represent the temperatures at which the transition is of first order. The inset in panel (a) shows effective exchange field $\bar{h}$ (full line) and the thin film approximation~\eqref{eq:h-vs-d-diffusive} (dashed line)  as a function of the thickness of the S layer for $h_{\rm ex}'=0.5\Delta_0\xi_S$.
    }
\end{figure}

The thickness dependence of the average exchange field \eqref{eq:spin-splitting-ballistic} and \eqref{eq:h-vs-d-diffusive} implies there is a critical thickness $d_c$ of the S film, defined by $\bar{h}(d_c)=h_c$, below which the film is not superconducting. Prediction from a semiclassical model \cite{hijano_coexistence_2021} is shown in Fig.~\ref{fig:phase_diagram}(a) for the critical interfacial exchange field, and in Fig.~\ref{fig:phase_diagram}(b) for the critical thickness. Inset of Fig.~\ref{fig:phase_diagram}(a) shows $\bar{h}(d)$, which is well approximated by Eq.~\eqref{eq:h-vs-d-diffusive} for thicknesses $d$ smaller than the coherence length $\xi_S$. For $d\gtrsim3\xi_S$, it becomes weak enough allowing superconductivity at zero temperature for any value of the interfacial field.

The magnetic properties of the FI also have important consequences for the physics.
Generally, the form of the magnetic hysteresis is determined by them, and magnetic details of the interface can matter
for the temperature dependence. Additionally, for bilayers with lateral dimensions surpassing the domain sizes in the FI, spatial variations in the exchange field introduce complexities that must be considered\cite{anderson1959spin,khusainov2002,aladyshkin2003domain,yang2004domain,strambini_revealing_2017,hijano_coexistence_2021,aikebaier2019superconductivity}. 
Moreover, the magnetic fringe fields of the FI lead to orbital depairing in the superconductor, generally suppressing
superconductivity and rounding sharp superconducting features in its density of states.

There are numerous additional factors that influence the experimental outcomes.
The physics is very sensitive to the interface quality:
on the one hand, conduction electrons that scatter back before interacting with localized FI magnetic moments, do not
acquire magnetic phase shifts or spin splitting. On the other hand, e.g. subgap bound states at the interface are predicted also to enhance the spin mixing. \cite{silaev2020,ahari2021} 
Hence, the behavior of the magnetic proximity effect is sensitive to the materials and growth conditions influencing the interface physics.
We return to these points in Sec.~\ref{subsec:materialspecific} and Sec.~\ref{sec:materials}.

\subsubsection{Interface transport model}
\label{subsec.S/FI transport}

The interplay of spin-splitting and spin-filtering in FI/S structures {is predicted to} lead to electron-hole symmetry breaking \cite{ozaeta_predicted_2014,machon2013,bergeret2018colloquium,heikkila2019thermal}. Namely, in Eq.~\eqref{eq:I(V)}, due to the combination of spin-dependence of the conductance ($G_\uparrow\ne{}G_\downarrow$), and the spin-dependent asymmetry of the density of states ($N_\sigma(-E)=N_{-\sigma}(E)\ne{}N_\sigma(E)$), it is possible to have $I(-V)\ne{}-I(V)$. These features enable both thermoelectric effects and non-reciprocal charge transport. 

\begin{figure}[t]
  \centering
  \includegraphics[width=0.8 \columnwidth]{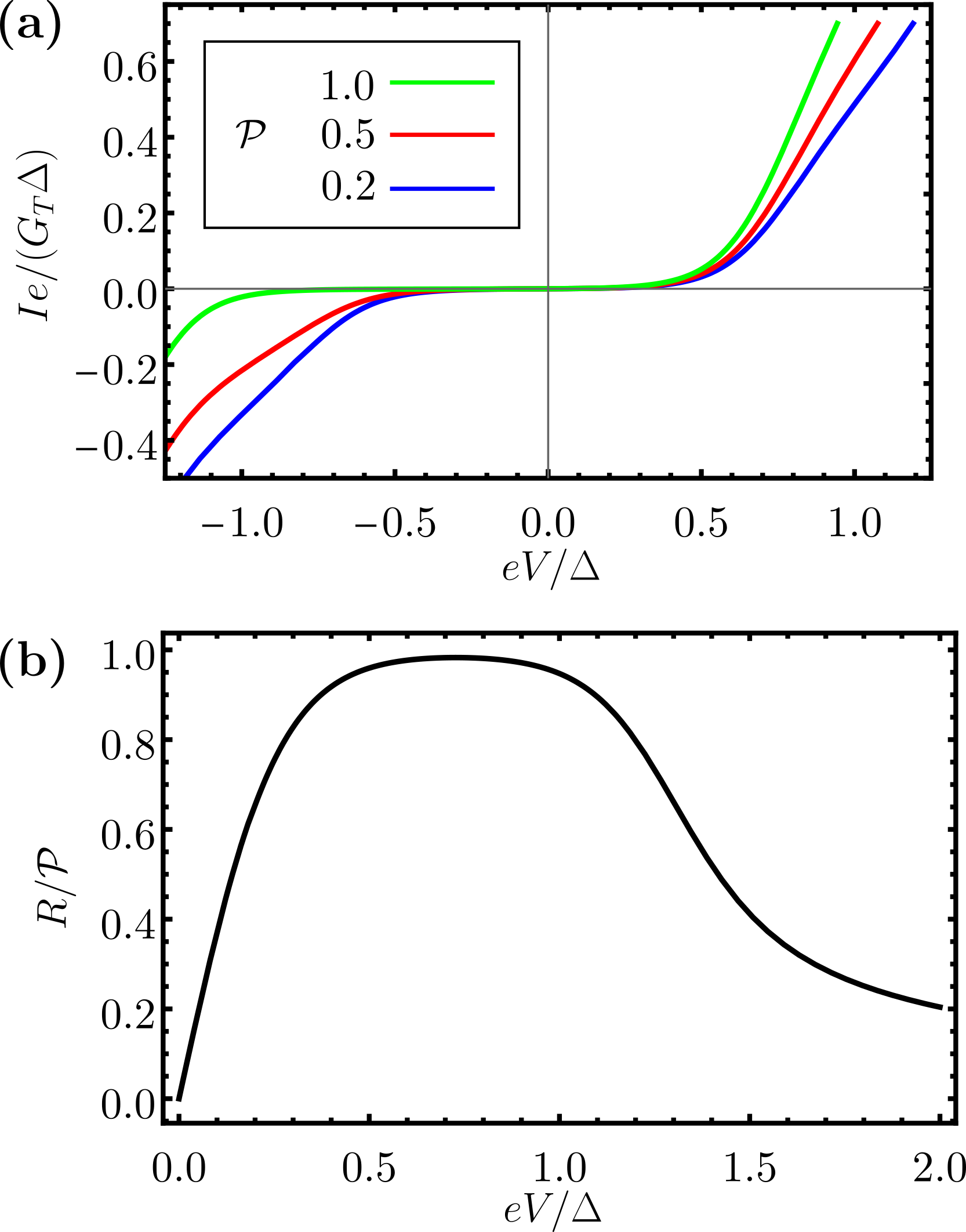}
  \caption{
    \label{fig:FigIV}
    (a) $I(V)$ curve of a N/FI/S junction for different values of the polarization $\mathcal{P}$. (b) Rectification coefficient $R=I_{\rm sym}/I_{\rm asym}$. Parameters used in the plots: $h=0.3 \Delta$, $T_N=T_S=0.1 \Delta$, $\Gamma=10^{-3} \Delta$. }
\end{figure}

\paragraph{{Thermoelectricity.}}

{For tunnel junctions, main features of spin-split superconductor thermoelectric effects are expected to follow Eq.~\eqref{eq:I(V)}.
Experiments \cite{kolenda_observation_2016,kolenda_thermoelectric_2017} indeed match its predictions well.}
For small temperature differences between the normal and superconducting sides of the tunnel junction, $\Delta T = T_S - T_N\ll T$, low voltages $eV< \Delta-h$, and low temperatures $k_B T\ll \Delta, h$,  Eq.~\eqref{eq:I(V)} reduces to \cite{ozaeta_predicted_2014, ilic_current_2022}
\begin{multline}
I(V,\Delta T)=-\alpha \frac{\Delta T}{T}+I_S \left ( e^{eV/k_B T}-1\right) \\
+ I_S (\mathcal{P}-1)\left( \cosh \frac{eV}{k_BT}-1\right).
\label{Eq:ThermRect}
\end{multline}
The first term describes the thermoelectric effect. Here, 
$\mathcal{P}=(G_\uparrow-G_\downarrow)/(G_\uparrow+G_\downarrow)$ is the spin polarization of the tunnel junction, and $\alpha$ is the thermoelectric coefficient, which in the ideal case~\eqref{eq:DOS-splitting} has the value $\alpha\approx \mathcal{P} (G_T/e) \sqrt{2\pi \tilde{\Delta}}e^{-\tilde{\Delta}}[\Delta \sinh (\tilde{h})-h \cosh{\tilde{h}}]$, \cite{ozaeta_predicted_2014} where $\tilde{\Delta}=\Delta/k_B T$ and $\tilde{h}=h/k_BT$. Moreover, $I_S=G_T/e K_1 (\tilde{\Delta})e^{\tilde{h}}$, where $K_1$ is a modified Bessel function.

Thermoelectric applications generally require, 
{in addition to the description of the electrical circuit where the thermoelectric element is embedded, also
modeling the induced temperature difference $\Delta T$. Generally, it depends on the various possible thermal conductance channels. Tunneling of electrons is one of these} and can also be described in the above approach, \cite{bergeret2018colloquium,heikkila2019thermal} but it is often not the most significant contribution. In particular, the electron-phonon coupling often dominates in a wide temperature range, see Refs.~\cite{giazotto_opportunities_2006,heikkila2019thermal} for a review. Generally, all heat conductances are exponentially suppressed in the superconducting state, due to the presence of the superconducting gap. This is why, in the presence of radiation with frequency larger than the gap, superconductors tend to heat up more than normal metals. A combined electrothermal theory for S/FI junction detector applications is discussed in more detail in Sec.~\ref{sec5}.

\paragraph{{Rectification.}}

{The second and the third term in  Eq.~\eqref{Eq:ThermRect} describe the non-reciprocal charge transport in the form of a non-ideal Shockley diode equation. Namely, the second term is the ideal Shockley contribution, while the third term describes deviations from ideal behavior, and vanishes at perfect polarization $\mathcal{P}=1$. The I-V characteristic is shown in Fig.~\ref{fig:FigIV}(a).}
The rectification can be characterized by decomposing the current into antisymmetric $I_{\rm asym}=\frac{1}{2}|I(V)-I(-V)|$ and symmetric $I_{\rm sym}=\frac{1}{2}|I(V)+I(-V)|$ parts in voltage, and defining the rectification coefficient
\begin{equation}
\mathcal{R}=\frac{I_{\rm sym}}{I_{\rm asym}}.
\end{equation}
The maximum of $\mathcal{R}(V)$ is determined by the spin-polarization of the junction, as shown in Fig.~\ref{fig:FigIV}(b), which provides a way to estimate the polarization parameter from the experimental $I(V)$ curves. 

In the presence of an AC voltage $V_{ac}$, N/FI/S junctions produce a rectified DC current, which at high frequencies and low temperatures ($\hbar \omega\gg \Delta, h, eV_{ac}$ and $k_B T \ll \Delta-h$) has the form \cite{ilic_current_2022}
\begin{equation}
I_{rc}=\frac{\mathcal{P}G_T}{2e}\frac{e^2 V_{ac}^2}{\hbar^2 \omega^2}h.
\label{Eq:AC}
\end{equation}
Note that the sign of the rectified current is determined by the relative sign of $h$ and $\mathcal{P}$, namely, $\text{sign}(I_{rc})=\text{sign}(h \mathcal{P})$. Similarly, for the current produced by the thermoelectric effect, $I_{th}=-\alpha \frac{\Delta T}{T}$, we have $\text{sign}(I_{th})=-\text{sign}(\Delta T \alpha)=-\text{sign}(\Delta T \mathcal{P} h)$. Therefore, $I_{th}$ and $I_{rc}$, have the same sign if $\Delta T<0$, meaning that the normal side of the junction is at a higher temperature. On the other hand, if the superconducting side of the junction is at a higher temperature ($\Delta T>0$), the two currents have opposite signs, so the rectification and thermoelectric effects compete. Therefore, to produce the maximal signal using FI/S junctions in e.g. radiation detectors, special care must be taken to optimize the ratio of rectification and thermoelectric effects, which depends on material-specific parameters and junction geometry.

The results above assume ideal I-V characteristics. As noted in Sec.~\ref{subsec:SFI}, several sources of nonideality can be phenomenologically described by introducing the Dynes parameter $\Gamma$ in the superconducting density of states. It enables subgap conduction channels, which can provide significant contributions at very low temperatures when $\frac{\Gamma}{\Delta} \gtrsim \frac{\Delta}{k_B T} e^{-\Delta/k_BT}$. In particular, the correction to the tunneling current due to small $\Gamma$ at $k_B T, h, eV \ll \Delta$ is \cite{ilic_current_2022}
\begin{equation}
\delta I=G_T \frac{\Gamma V}{\Delta}\bigg( 1+\frac{3}{2} \frac{e V h}{\Delta^2}\bigg),
\label{eq.DeltaI}
\end{equation}
yielding the rectification coefficient of $R\approx 3 \mathcal{P} eV h/(2\Delta^2)$.
Similarly, nonzero $\Gamma$ also affects the thermoelectric response. At low temperatures $k_B T \ll \Delta-h$ and for small $\Gamma$ the thermoelectric coefficient obtains an extra contribution 
\begin{equation}
    \delta\alpha=\frac{G_T h \pi^2 k_B^2 T^2 \Delta^2 \Gamma}{e(\Delta^2-h^2)^{5/2}}.
    \label{eq.DeltaAlpha}
\end{equation}
On the other hand, the thermoelectric voltage (or Seebeck coefficient), represented as $-\alpha k_B \Delta T/(eG)$, generally decreases due to $G$ increasing more than $\alpha$ \cite{ozaeta_predicted_2014}. Besides the Dynes parameter, somewhat similar effects on the thermoelectric coefficients are provided by spin-flip scattering as described in \cite{rezaei2018}.

\paragraph{{Detector yield.}}

The efficiency of conversion of absorbed radiation
can be characterized by the quantum yield, i.e.,  the number $n_{\rm th}(\omega)$ of electrons transported
across the junction per absorbed photon at frequency $\omega$.
For steady-state thermoelectric conversion, the estimation can be made using Eq.~\eqref{Eq:ThermRect} and a heat balance model
\begin{equation}
    a_{S/N} P_\gamma = G_{\rm th}^{S/N} (T_S-T_{\rm bath})\pm G_{\rm th}(T_S-T_N).
\end{equation}
This describes fractions $a_S$ and $a_N$ of the absorbed power $P_\gamma=\hbar\omega/\delta t$
distributed on the S and N sides in time $\delta t$, and relaxing via junction and parasitic thermal conductances $G_{\rm th}$ and  $G_{\rm th}^{S/N}$. Now, assume that $k_B T\ll\Delta,h$ so that $G_{\rm th}^S\ll{}G_{\rm th}\ll{}G_{\rm th}^N$, \cite{giazotto_opportunities_2006,heikkila2019thermal} and that $\hbar\omega\gtrsim2\Delta$
so that absorption $a_S$ is non-negligible. Using the low-temperature values of the
junction transport coefficients,
\begin{equation}
    n_{\rm th} 
    = I_{\rm th} \delta t/e
    \approx a_S \mathcal{P} \frac{\hbar \omega}{\Delta-h} \sgn(h),
\end{equation}
for $\Gamma\lesssim{}e^{-\Delta/k_BT}\Delta^2/h$. In the opposite limit
of large $\Gamma$, $n_{\rm th} \approx 3 a_S \mathcal{P} \hbar \omega h/(\Delta^2 -h^2)$
which is smaller by only a factor of $h/\Delta$.  Hence, absorption of a single
photon fairly generally transports $n_{\rm th}>1$ electrons across the junction.

A similar estimate can be made for the rectification effect, assuming that fraction $a>0$ of the incoming power $P_\gamma$ generates an AC voltage across the N/FI/S junction, $a P_\gamma=V_{ac}^2/(2\tilde{R})$. Here $\tilde{R}=R/(1+R G_T)$ is the parallel resistance of the junction and a radiation absorber of resistance $R$. Using Eq.~\eqref{Eq:AC},
\begin{equation}
  n_{\rm rc}=I_{\rm rc}\delta t/e=a \mathcal{P}\frac{h}{\hbar \omega} G_T \tilde{R}.
\end{equation}
Here $n_{rc}<1$ at $\hbar\omega\gtrsim\Delta$, indicating that the thermoelectric
effect can dominate AC rectification, and lead to better device performance.

\subsubsection{Material specific implementation}
\label{subsec:materialspecific}

The results above highlight that for practical purposes it is desirable to find suitable superconductor/ferromagnet combinations. Thermoelectricity and non-reciprocal effects are maximized by having, on one hand, a good spin-filter efficiency and, on the other hand, a sizable spin-splitting of the superconductor.  {Both of these effects} depend crucially on interfacial parameters. As discussed in the next section,  very good {experimental} control over  spin split superconductivity has been achieved in EuS/Al systems \cite{meservey_spin-polarized_1994,hao_spin-filter_1990,hao1991,hao_magnetic_2010,strambini_revealing_2017,de_simoni_toward_2018,strambini2022superconducting}.  
Crucial for the splitting is the size of the imaginary part of the mixing conductance, as explained in section \ref{subsec:SFI}. 
This value can be notably large at EuS/metal interfaces, as demonstrated in spin Hall magnetoresistance experiments on EuS/Pt systems \cite{gomez2020strong}, which found
$h_{\rm ex}'\approx{}0.8\,\mathrm{meV\,nm}$, as extracted from measured $\Im G_{\uparrow\downarrow}$ using Eq.~\eqref{eq:Gupdownmod}. This value corresponds to the superconducting gap $\Delta_{\mathrm{Al}}$ of aluminum with a film thickness of $d\approx4 \,\mathrm{nm}$.
In EuS/Al the induced effective exchange field can range from values much smaller than $\Delta_{\mathrm{Al}}$, up to such large values that superconductivity is fully suppressed, as observed in \cite{li:2013gh}.  

Recent experiments also suggest a finite induced  exchange field in   YIG/NbN \cite{rogdakis2019spin} and 
YIG/Nb \cite{jeon2020giant,jeon2022zero}, even though 
the spin-mixing conductance of YIG/metal interface is dominated by the real part and not the imaginary part, in contrast to EuS based structures \cite{gomez2020strong}. 

\hyphenation{chalco-gen-ides}

With respect to spin-selective transport, Eu-chalcogenides serve as highly efficient spin-filtering barriers.  Efficiencies larger than 95\% have been achieved in  EuSe barriers, whereas 80\% spin-filter efficiency has been achieved in EuS barriers \cite{moodera_phenomena_2007}.  
However, a fundamental challenge arises when using these magnetic barriers in several of the devices discussed herein: achieving high spin filter efficiency necessitates thick barriers, which subsequently exhibit substantial resistances.  "One way to circumvent this issue is to utilize a ferromagnet as the normal metal electrode in lieu of a magnetic barrier. \cite{kolenda_thermoelectric_2017,strambini2022superconducting}.   Indeed the thermoelectric effect has been observed in EuS/Al/Co \cite{strambini2022superconducting},  EuS/Al/Fe \cite{kolenda_thermoelectric_2017}, and in  Al/Fe tunnel junctions \cite{kolenda_observation_2016}. In the latter, an external magnetic field was applied to provide the spin-splitting in the superconductor. 
 A spin filtering efficiency of 75\% was reported in NbN/GdN/NbN Josephson junctions \cite{senapati_spin-filter_2011}, and a sizable interfacial exchange coupling {was seen} in Nb/GdN samples \cite{cascales2019switchable}.

\begin{table*}
\caption{ Summary of reported values for spin filtering efficiency (barrier polarization, $\mathcal{P}$) and exchange splittings ($\bar{h}$) across different experimental studies. Color coding indicates {\color{red} ferromagnets} and {\color{blue} spin-split superconductors}. \label{table:FIS} }%
\begin{indented}\item[]\begin{tabular}{@{}llll}
\br
      Material combination & Barrier polarization $\mathcal{P}$ & Exchange splitting $\bar{h}$ @ applied field $B$ & Reference
      \\
      \mr
      \textcolor{red}{EuO}/\textcolor{blue}{Al}/AlO$_3$/Al & no spin-filter barrier &1 T @ 0.1 T \dots 1.73 T @ 0.4 T & \cite{tedrow_spin-polarized_1986} \\
  Au/\textcolor{red}{EuS}/\textcolor{blue}{Al} & 0.8 &1.6 T @ 0 T & \cite{moodera_electron-spin_1988} \\
\textcolor{blue}{Al}/\textcolor{red}{EuS}/Al & 0.6-0.85 & 1.9-2.6 T @ 0 T & \cite{hao_spin-filter_1990} \\
Ag/\textcolor{red}{EuSe}/\textcolor{blue}{Al} & $>$ 0.97 & 0 T @ 0 T & \cite{moodera_variation_1993} \\
  \textcolor{red}{EuSe}/\textcolor{blue}{Al}/AlO$_3$/Ag & no spin-filter barrier & 4 T @ 0.6 T & \cite{moodera_variation_1993} \\
  \textcolor{blue}{NbN}/\textcolor{red}{GdN}/\textcolor{blue}{NbN} & 0.75 & $-$ & \cite{senapati_spin-filter_2011}  \\
     \textcolor{blue}{NbN}/\textcolor{red}{GdN}/TiN & 0.97 & 1.4 T @ 0 T & \cite{pal2015large}  \\
    \textcolor{red}{EuS}/\textcolor{blue}{Al}/\textcolor{red}{Fe} & 0.15-0.20 &  1.5 T @ 0.4 T & \cite{kolenda_thermoelectric_2017} \\
    Cu/\textcolor{red}{EuS}/\textcolor{blue}{Al} & 0.48 & 1.83 T @ 0.1 T & \cite{strambini2022superconducting}\\
    \textcolor{red}{EuS}/\textcolor{blue}{Al}/AlO$_x$/\textcolor{red}{Co}  & 0.3 & 1.67 T @ 0 T & \cite{strambini2022superconducting}\\
    \textcolor{red}{EuS}/Al/AlO$_x$/Al & no spin-filter barrier & 2.52 T @ 0 T & \cite{hijano_coexistence_2021} \\
    \br
    \end{tabular}\end{indented}
\end{table*}

 Table \ref{table:FIS} summarizes various experimental studies on FI/S structures, providing values for the spin filtering efficiency and the spin-splitting field.

\subsection{Materials and device preparation}
\label{sec:materials}

For cryogenic thermoelectric detectors and non-reciprocal electronics that utilize spin-polarized tunneling junctions with exchange-split superconducting electrodes, there are application-dependent considerations that for example require maximizing the amplitude of the thermoelectric effect, which depends on the splitting of DoS, or setting limits on the tunnel junction resistance to achieve rapid readout. {This implies having good control on the properties of the devices constituent material as well as the growth parameters of the heterostructures}. In this section, we focus on the elements concerning the interface between a superconducting thin film and a magnetic insulator or semiconductor that creates a spin-split density of states. Application-dependent issues are discussed in the corresponding sections. In the early works of Meservey and Tedrow, many different material combinations were tested, including chalcogenides of  rare-earth elements as well as very thin metallic layers interfaced with Al and V 
layers (for a review see \cite{meservey_spin-polarized_1994}).
The effect of the exchange splitting is usually  explored via tunneling spectroscopy.  %
Spin splittings corresponding to effective magnetic fields of the order of several Tesla  have been achieved, in particular using ferromagnetic insulators.
The splitting decreases with the increase in thickness of the superconducting film in close accordance with the predictions of the de Gennes' theory \cite{degennes1966-cfs}. 
In the case of metallic magnetic layers adjacent to superconducting films the situation is more subtle since Cooper pairs can leak into the magnetic region leading, in general,  to a suppression of superconductivity. For this reason, in order to obtain a good splitting, it is more convenient to use magnetic insulators. %
Many different ferromagnetic insulators (FI), including transition metal oxides or rare earth oxides, such as gadolinium gallium garnet (Gd$_3$Ga$_5$O$_{12}$) and yttrium iron garnet (Y$_3$Fe$_5$O$_{12}$) were tested. However, EuS was the first FI where spin filtering was determined \cite{moodera_electron-spin_1988}, which has implications in magnetic tunnel junctions and spin valves \cite{de_simoni_toward_2018}, as well as in superconducting electronics, and radiation sensors \cite{hao_spin-filter_1990,kolenda_thermoelectric_2017,miao_magnetoresistance_2009,moodera_electron-spin_1988,wolf_fabrication_2014,hijano_coexistence_2021}.

To obtain reliable and reproducible amplitudes of the exchange splitting, it has been demonstrated that the quality of the interfaces is crucial \cite{hao_spin-filter_1990}. Introducing a thin insulating layer between the magnetic and superconducting films results in a complete quenching of the exchange interaction \cite{hao_spin-filter_1990}. Moreover, in many examples where lithography-based nano-fabrication was used to create devices with exchange split DOS, especially based on the combination of Al and EuS \cite{xiong_spin-resolved_2011,wolf_spin-polarized_2014}, the exposure of EuS to atmosphere before the deposition of the superconductor prevents maintaining the exchange splitting without external magnetic field even after the initial saturation of the magnetization below the Curie temperature of the EuS. 
These findings suggest that the EuS was partially oxidized during the lithographic nanofabrication, which causes formation on the interface of the paramagnetic Eu-based compound with zero magnetic remanence.
In contrast, in several references of devices prepared by Moodera and coworkers, a zero-field exchange split DoS was demonstrated in the absence of any external magnetic field \cite{moodera_exchange-induced_1989,strambini_revealing_2017}. {The explanation lies in the method of device fabrication, which is \textit{in-situ} shadow mask evaporation. In the early works of Meservey and Tedrow, the crossed-wire tunneling junctions used in tunneling spectroscopy experiments were created by subsequent deposition of the FI and S materials using different shadow masks in a high vacuum environment
 (1x10\textsuperscript{-8}~mbar) and preventing the exposition of the layers to the air \cite{tedrow_spin-polarized_1986,hao_spin-filter_1990}. More recently, it has been shown that the EuS sublimation temperature also plays a significant role, with a trend toward Eu-rich compounds at higher sublimation temperatures \cite{gonzalez_Orellana_2023}. This modification to the chemical composition is mirrored in the magnetic properties, with an increase in magnetic moment and Curie temperature. Therefore, the need for precise control of the composition and cleanliness of the interface between the superconductor and the magnetic layer places constraints on the fabrication techniques for tunneling junctions.}

To illustrate the growth of the required heterostructures for fabricating cryogenic thermoelectric detectors or non-reciprocal electronics, in the subsequent paragraph, we detail the shadow mask fabrication procedure of the EuS/Al-base devices described in Ref.~\cite{hijano_coexistence_2021}, which was similar to the one summarized long time ago in the review by  Meservey and Tedrow\cite{meservey_spin-polarized_1994}. To complement it, we discuss those aspects relevant to understand and control the device performance. These example potentially serve as a foundational elements for a SFTED.

\begin{figure*}[t]
    \centering
    \includegraphics[width=0.8\textwidth]{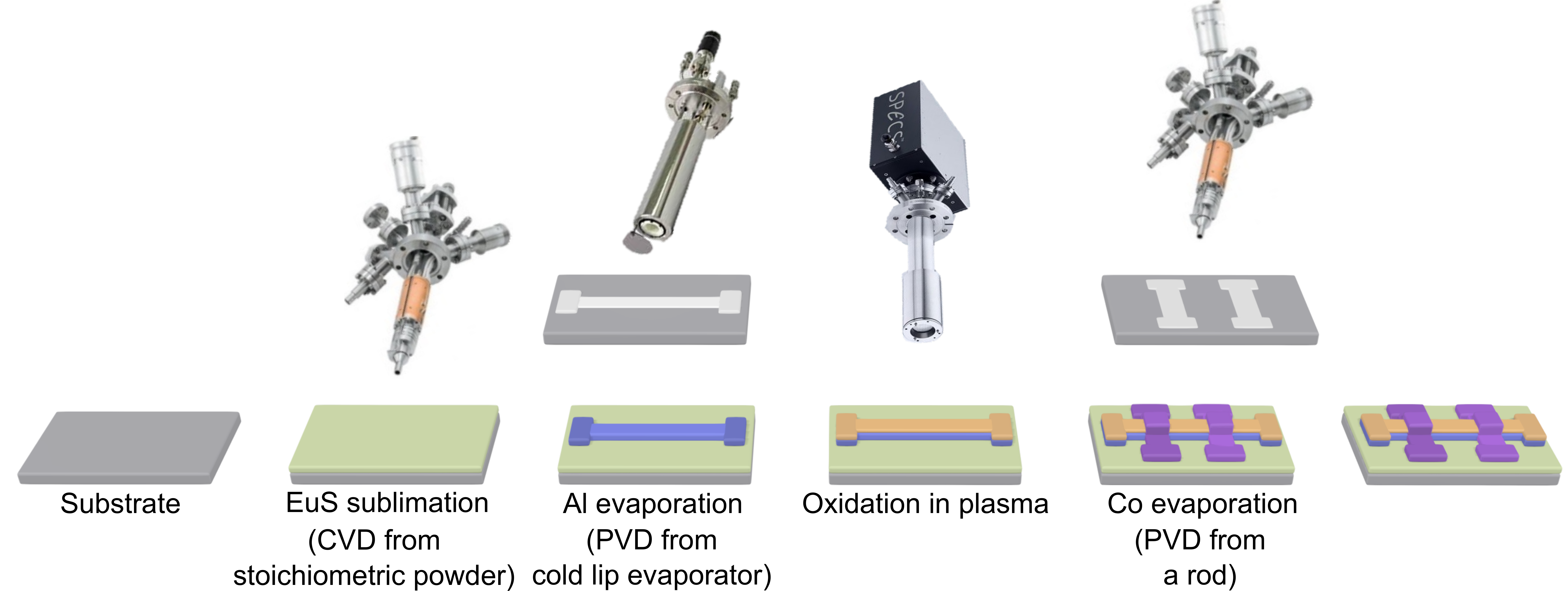}
    \caption{Schematic representation of the different steps involved in the hybrid heterostructure growth.}
    \label{Fab_overview}
\end{figure*}

\subsubsection{Device Fabrication}

{As we mentioned before, in Ref. \cite{hijano_coexistence_2021} the procedure to fabricate X/AlO$_x$/Al/EuS tunnel junctions was essentially shadow mask fabrication. Here, X denotes a metallic layer, either superconducting or magnetic.} The sequence is schematically represented in  Fig.~\ref{Fab_overview}, and it includes: {i)} the deposition of the magnetic EuS on the insulating substrate via electron-beam (e-beam) sublimation from stoichiometric powder; {ii)} the growing of the superconducting Al electrode through the shadow mask; {iii)} the formation of the tunneling barrier via the oxidation in the oxygen plasma; {iv)} the deposition of the counter electrodes X through another shadow mask; and {v)} finally deposition of the capping layer to protect the devices from the atmosphere. All the process is done without breaking the vacuum at any time, and therefore preventing the undesirable oxidation or contamination of the layers and interfaces.

{The substrates for fabricating cryogenic thermoelectric detectors or non-reciprocal electronics must be good insulators, such as high-quality polished silica or SiN. Thin film of EuS layer is deposited as a continuous film by employing a Knudsen cell (Tungsten crucible) filled with commercially available stoichiometric EuS powder, as Smoes and coworkers demonstrated in 1977 \cite{Smoes_1977}. During the sublimation process, 
the crucible temperature {can be} estimated via the power vs temperature calibration curve and the growth rate {is} calibrated using a quartz microbalance, or monitored via a flux monitor. This is relevant for sample preparation because EuS powder decomposes during evaporation  and the ratio Eu:S in the gas phase varies with increasing temperature, which affects its magnetic behaviour \cite{Smoes_1977}. }
{Superconducting and magnetic metallic wires, Al and Co in example of Fig.~\ref{Fab_overview} must be deposited over the EuS layer in the same vacuum chamber, i.e. without breaking the vacuum, in order to avoid the oxidation of the EuS surface. There are several methods to deposit aluminum layers in vacuum environments, being the thermal evaporation used in Ref. \cite{hijano_coexistence_2021} one of the most convenient, because of its reproducibility, its cleanness and its precise control of film thickness and composition. }
{Although for the SFTED, the second metallic  electrode needs to be a ferromagnet, as for example Co, for the characterization of the spin-splitting induced by the EuS in the Al wire, it is convenient to first grow an Al/AlO$_x$/Al/EuS structure \cite{moodera_electron-spin_1988,hao_spin-filter_1990,strambini_revealing_2017,hijano_coexistence_2021}.}

{Regarding oxide barrier fabrication, the conventional method to grow the AlOx barrier is thermal oxidation, where aluminum is directly exposed to air (or pure molecular oxygen), resulting in many instances in amorphous layers that lead to inhomogeneous interfaces \cite{Shih2006}. Moreover this method is self-limited in the barrier thickness, being only possible to grow barriers of around 2 nm \cite{Baran}. Among alternative methods to produce tunnel barriers, the most popular is the oxygen glow discharge  plasma oxidation that employs atomic oxygen \cite{Kuiper2001} instead of molecular one to produce much thicker barriers in shorter time. However, this is an aggressive technique because the oxygen atoms that arrive to the sample have high kinetic energy and can  create defects in the AlOx layer. This drawbacks can be overcome by Inductively Coupled Plasma sources, which produce barriers with very low amount of pinholes and high reproducibility \cite{Ando_2002, Gonzalez_2023}.}

{Typical tunneling spectroscopy measurements of the Al/AlOx/Al/EuS junctions shown in  Fig.~\ref{Fig._Inital-charac} revealed the spin-split DoS of the bottom Al electrode even in the absence of an external magnetic field. }
The characteristic splitting of the differential conductance ($dI/dV$) curves becomes more pronounced when the magnetization of the sample is saturated, and remains almost intact when the field is reduced back to zero. 
{ Work \cite{gonzalez_Orellana_2023} discussed the possibility of a systematic control of the exchange coupling strength via tuning of the parameters of sublimation of EuS. The control of the effective exchange field is essential to optimize the SFTED.}

\subsubsection{EuS properties}
Early works aimed at the growth of thin films of EuS unveiled a strong variation in their chemical composition associated with different deposition conditions~\cite{Kohne_1980, guilaran_observation_2003}. Since the evaporation of sulfur, which is required for Molecular Beam Epitaxy (MBE), is technically challenging, the preferred technique for growing EuS is sublimation from stoichiometric powder. Detailed study of the thermodynamics of this process showed that solid EuS decomposes due to the heating and the proportion of Eu to S in the gas phase changes slowly with increasing temperature from nearly 1:1 ratio at 1400~K to the Eu-reach mixture at temperatures above 2000~K~\cite{Smoes_1977}. Usually, the temperature of sublimation is kept high to obtain higher growth rate. Excess of Eu in the deposited films is compensated by means of heating of the substrate. For instance, it has been demonstrated that increasing the temperature to 630~K results in near-stoichiometric films, while cooling the substrate promotes the growth of Eu-rich films~\cite{Kohne_1980, guilaran_observation_2003, Keller_2002}. 
In the pioneering works of Moodera's group on  the fabrication of the exchange-coupled Al/EuS bilayers, EuS (and Al) are usually deposited on the substrates cooled down with liquid nitrogen because it results in higher magnetic remanence (more rectangular hysteresis loops), which is an advantage for the applications that require persistence of the exchange-coupling between the EuS and Al in zero magnetic field. {A recent work, Ref.~\cite{gonzalez_Orellana_2023}, has shown that deposition of the EuS on the substrates kept at room temperature or cooled with liquid nitrogen has to be accompanied by the control of the sublimation temperature in order to limit the exchange coupling strength in the Al/EuS - based devices.} It was found that the interaction gets stronger with the increase in Eu concentration. It quickly surpasses the critical value, quenching the superconductivity of the thin Al films, once the Eu/S ratio becomes higher than unity~\cite{gonzalez_Orellana_2023}.  {Control of the power} of the e-beam evaporators allows for precise manipulation of the EuS sublimation temperature, which opens up a way to the tuning of the exchange coupling.
{For instance, in the series of samples listed in Table~\ref{tab:table2}, } {effective exchange field variations of up to a factor of three are obtained by tuning the layer thicknesses and/or the growth temperatures.}
 
{The capability to control the exchange coupling strength in the Al/EuS bilayers enables the validation of the theoretical description of the proximity effect for hybrid S/FI systems presented in Section~\ref{subsec:SFI}.}%
{Furthermore, by tuning the exchange coupling strength to be near the critical value, it becomes possible to investigate the magnetization reversal process in EuS films across a temperature range below 1.2 $K$. This range is typically inaccessible for the majority of magnetic measurement systems~\cite{hijano_coexistence_2021}.} 
{Figure~\ref{Fig._Inital-charac}~(c) presents typical results for tunneling spectroscopy of the Al/AlO$_x$/Al/EuS tunnel junction. In this spectrum, the characteristic splitting induced by the exchange interaction between the superconductor and the magnetic insulator is observable within a narrow range of external magnetic fields.}  This behaviour was interpreted in Ref. \cite{hijano_coexistence_2021} as decreasing of the average effective field in the Al/EuS bilayer close to the coercive field and quenching of the superconductivity in the Al wire interfaced with EuS, once the applied field saturates the magnetization of the magnetic layer. Thorough measurements of the transport properties of the Al wire coupled to EuS showed that magnetization reversal causes step-wise changes of the resistance. This indicates local quenching of the superconductivity. Phase diagram built for the range of temperatures from 30~mK to 1.2~K and within the $\lvert B \rvert \lesssim$~50~mT allowed for tracking a variation of the coercive field with temperature. This unveiled a number of critical temperatures that we attributed to the presence of small superparamagnetic particles of EuS~\cite{hijano_coexistence_2021}. These findings were correlated with results of magnetic characterization by means of XAS/XMCD spectroscopy, magnetometry and low-temperature ferromagnetic resonance (FMR)~\cite{Aguilar_2023, gonzalez_Orellana_2023}. For instance, it was observed that sulfur-rich and near stoichiometric polycristalline EuS films with moderate exchange coupling strengths show very broad and shallow peaks of FMR, meanwhile the Eu-rich films possess sharper resonance but their coupling to thin Al films causes quenching of the superconductivity.                   
Besides the preparation of polycrystalline films, growth of textured and single crystal epitaxial EuS layers was reported~\cite{guilaran_observation_2003, Demokritov_1996, Liu_acsami_2020, Liu_nanolett_2020, katmis_high-temperature_2016, Goschew_2017}. The temperature of the substrate and the density of dislocations were crucial to controlling the stoichiometry of  EuS~\cite{guilaran_observation_2003, Demokritov_1996}. Small lattice mismatch and suitable range of the substrate thermal stability made possible the growth of very high-quality epitaxial EuS/InAs(001) films~\cite{Liu_acsami_2020}. These layers were employed to rigorously study their magnetic and electronic properties and to grow the fully epitaxial Al/EuS/InAs(001) heterostructures~\cite{Liu_nanolett_2020}. Together with results obtained in the GdN/NbN system~\cite{linder2015superconducting}, these works make up a solid base for understanding the exchange coupling phenomena in the epitaxial S/FI layers. 

An alternative approach to the fabrication of heterostructures, which may overcome the challenges associated with traditional growth methods, involves utilizing two-dimensional (2D) materials. Van der Waals stacking of magnetic insulator and superconductor layers, as reported in several recent studies, has produced high-quality samples exhibiting a diverse array of intriguing properties. This showcases the significant potential of this technique \cite{Kezilebieke_nanolett_2022, Aikebaier_VdWheterostructures_2022}.   Inherently, 2D materials exhibit a sharp interface between adjacent layers in their stacks. This permits the engineering of heterostructure interfaces with a level of precision only achievable with the most advanced Molecular Beam Epitaxy systems \cite{Pham_2022}.
Due to minimal chemical intermixing between layers, it is possible to combine distinct slabs of various 2D materials and even single atomic layers in the case of graphene or hexagonal boron nitride. This precision opens up a way to study the phenomena typically unobserved in bulkier heterostructures. For example, a rich assortment of strongly correlated electronic states is evident in twisted bilayer graphene \cite{Andrei2020}. On the other hand, sharp well-defined interfaces between dissimilar 2D materials enable engineering of interface-dependent effects, like exchange splitting of the electronic density of states in the superconductors \cite{Aikebaier_VdWheterostructures_2022}. For instance, a heterostructure that integrates an insulating layer of the ferromagnetic CrBr$_3$ and superconducting NbSe$_2$ showcases these capabilities \cite{Kezilebieke_nanolett_2022}. While some 2D materials exhibit strong reactivity and chemical instability, a possibility of obtaining single layers via exfoliation is widely accepted as a feature of the 2D materials, which facilitates building of the heterostructures by means of stacking of the layers via mechanical micromanipulations. Nevertheless, exfoliation produces flakes of variable thickness that requires evaluation and selection procedure incompatible with automatic fabrication processes. Therefore, technologies for growing macroscopic (wafer-size) single-layer films and the protocols of transferring of these films have being developed for a number of most relevant 2D materials \cite{Pham_2022,zhang_strategies_2021}. Such advancements pave the way for implementing 2D materials in the applications and creates incentives for further research activity in the field. However, several challenges still need to be addressed to fully leverage the benefits of this new group of materials. In terms of materials science, significant work is required to broaden the selection of materials available as large, high-quality wafer-size single layers. While progress is being made in developing growth processes for semiconducting and superconducting materials, exfoliation remains the most common method for obtaining single-layer magnetic 2D material\cite{Pham_2022,zhang_strategies_2021}. An alternative to wafer-scale production of 2D materials could involve the development of moderate-temperature growth techniques for 2D layers, which could be compatible with standard lithography-based nanofabrication \cite{bikaljevic_noncollinear_2021}. On the front of device physics and technology, significant efforts will be required to adapt fabrication processes to the new materials and address common challenges associated with implementing electronic devices, such as establishing electrical contacts to the 2D layers  \cite{Chhowalla2016, ZhengOhmic2021}. Specific challenges related to realizing superconducting electronic devices with non-reciprocal transport properties include determining the spin-filtering efficiency and the strength of spin splitting due to exchange coupling in heterostructures of 2D superconductors and 2D magnetic insulators. This topic requires further experimental investigation.

\subsection{Tunnel characteristics \label{sec4}}
In this section, we summarize the low-temperature characterization methods, which provide information on spin-splitting and polarization of the tunneling transport and how to improve it. 
We also discuss experiments on the thermoelectric and rectification effects in different junctions based on S/FI elements. 
\label{sec:tunnel-characteristics}

\subsubsection{Characterization and optimization of the exchange splitting }

\begin{table*}
\caption{\label{tab:table2}Overview of the investigated samples with varying layer thicknesses and growth temperatures.}
\begin{tabular}{cccccc}
\br
Samples & Structure & $\Delta_{1}$ ($\mu$eV) & $\Delta_{2}$ ($\mu$eV) &  $\Bar{h}$ ($\mu$eV) & $\Bar{h}/\Delta_{2}$ \\
\mr
S1   & Al (10nm)/ AlOx/ Al (12nm)/ EuS (19nm)/ silica                  &  254 & 131 & 68 & 0.5 \\
\mr
S2   & Al (10nm)/ AlOx/ Al (10nm)/ EuS (11.5nm)/ silica                &  & & 118 & \\
\mr
S3   & Al (12nm)/ AlOx/ Al (12nm)/ EuS (16.5nm)/ silica              &  200 & 275 & 150 & 0.57 \\
\mr
S4   & Al (12nm)/ AlOx/ Al (6nm)/ EuS (10nm)/ silica$^a$  &  & &$\Bar{h} > \Bar{h}_{cr}$ & \\
\br
\end{tabular}
\footnotesize{$^a$EuS and Al grown at cold $T$ (150 K)}
\end{table*}

The properties of the ferromagnetic superconductor generated with a FI/S bilayer are heavily influenced by factors such as operating temperatures and external magnetic fields as well as material characteristics including the quality of the interface between the two layers, the thicknesses of layers and the material selection for the FI and S layer~\cite{hijano_coexistence_2021}. Different materials combinations are discussed in Sec.~\ref{subsec:materialspecific}. 
In the most typical case of using Al as the superconductor, the critical temperature is relatively low ($T_C \sim 1-2$ K, depending on the film thickness), but also the spin-orbit interactions (SOI) are weak due to the low atomic mass of Al. In the presence of disorder, SOI leads to spin relaxation \cite{bergeret2018colloquium} and therefore the spin-splitting essential for non-reciprocal electronic properties of the tunnel junction is clearest in metals with weak SOI.
By contrast, the disadvantage of operating below 1~K
is not  significant for detection applications that typically need to operate at ultra-low temperatures to minimize thermal noise.

The FI that pairs optimally with Al is EuS due to a minimal lattice mismatch making ideal Al/EuS interfaces so that a strong ferromagnetic exchange interactions can be induced in the Al. Such interactions generate an effective Zeeman splitting ($\bar{h}$) in the DoS comparable to the Al superconducting gap ($ \Delta_0 \simeq 200$ $\mu $eV), i.e., equivalent to the Zeeman splitting induced by an external magnetic field of a few Tesla. For the fabrication of optimal FI/S bilayers two experimental parameters are crucial: the quality of the FI/S interface to induce strong exchange interactions in the S film and the thickness of the S that facilitates such interactions through the film thickness $d$. It has been demonstrated~\cite{hijano_coexistence_2021} that for $d\geq 3 \xi_S $ the induced $\bar{h}$ is almost negligible, determining the upper limit to the S thickness.
On the other hand, to maximize $\bar{h}$ the Al thicknesses ($d$) can be lowered~\cite{hao1991} down to a critical thickness ($d_c$) below which the ferromagnetic correlations increase up to the Chandrasekhar-Clogston limit and kill superconductivity as described in Sec.~\ref{subsec:SFI}. Using Eq.~\eqref{eq:h-vs-d-diffusive}, in thin layers such a lower limit can be estimated from the Zeeman splitting of a sample as:
\begin{equation}
    d_c =  \frac{\sqrt{2} d \bar{h}(d)}{\Delta_0} \simeq \frac{\sqrt{2} h_{\rm ex}'}{\Delta_0}.
\end{equation}

It is worth noting that from this equation, $d_c$ depends only on the intrinsic parameters $h_{\rm ex}'$ and $\Delta_0$. As described in Sec.~\ref{subsec:materialspecific}, for an Al/EuS bilayer with $\Delta_\mathrm{Al}\simeq 200$ $\mu$eV and $h_{\rm ex}'\simeq 0.8$~meV~nm it is possible to estimate $d_c \simeq 4$ nm. This limit has been numerically evaluated and demonstrated by comparing the tunneling spectroscopy of samples made with Al/EuS bilayers of different thicknesses~\cite{hijano_coexistence_2021}. 

Besides the direct magnetic proximity effect, external magnetic fields can also be exploited to enhance $\Bar{h}$. Interestingly, a large increase is typically observed even at small fields for which the simple additional Zeeman term would be negligible. This suggests a superparamagnetic behavior in EuS since the spins at the Al interface surface likely have a weak coupling with the bulk EuS. \cite{strambini_revealing_2017}. 
This effect has already been observed in early experiments~\cite{kolenda_thermoelectric_2017, strambini_revealing_2017, de_simoni_toward_2018} and in some samples can be strong enough to induce a S to N phase transition. 

\begin{figure*}[t]
    \centering
    \includegraphics[width=1.1\textwidth]{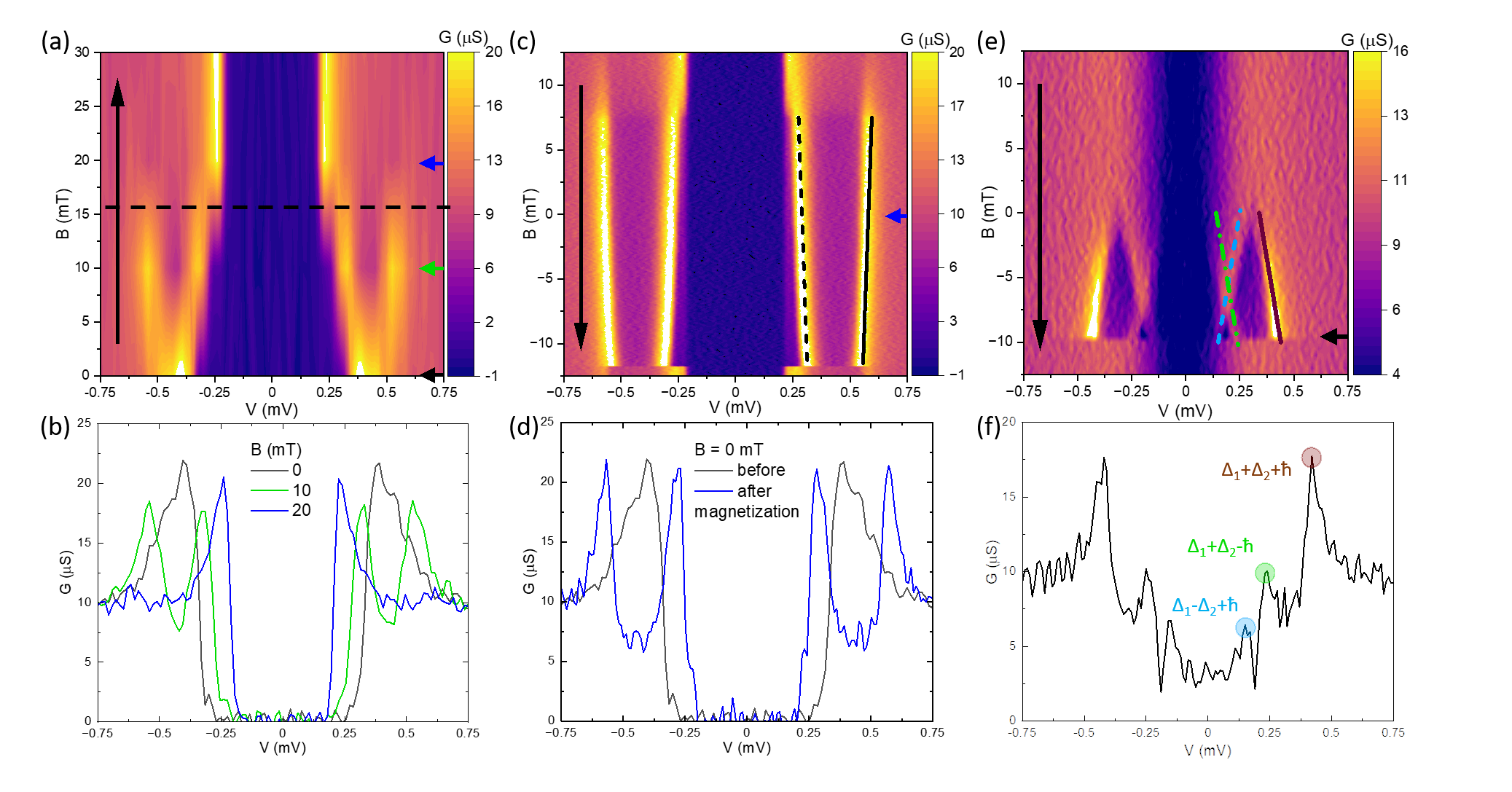}
    \caption{
    \textbf{Tunneling spectroscopy of sample S3-e.} 
    (a) \textbf{First magnetization of the EuS.} 
    Before the application of any magnetic field ($B=0$, black arrow), a clear $\Bar{h}$ is observed (broad peak around $V= \pm 0.4 mV$) in the tunneling conductance. It is a signature of the magnetization of the individual EuS grains not being aligned with respect to one another. %
    Sweeping from 0 to 30 mT said broad peak quickly splits into two (the magnetization directions of the individual EuS grains are being aligned) and then disappears above around 16 mT (dashed line) where the superconductivity of one Al layer is quenched (the bottom Al layer which is in contact with the EuS). Measurement done at $T=30$ mK.
    (b) Evolution of the peaks in (a) taken at $B=0$ (black), $10$ mT (green) and $20$ mT (blue), indicated in (a) by arrows.
    (c) \textbf{Magnetic field sweep of the tunneling spectroscopy at $T=30$ mK.} After the first polarization of the EuS layer, a spin-splitting in the tunneling spectroscopy is observed between around +8 mT and -12 mT. The split peaks move slightly towards each other throughout the sweep. Here, the surface spins of the EuS are dependent on and therewith behaving like the EuS grains which are responsible for the narrowing. Beyond mentioned applied magnetic field strengths, the spin-splitting disappears. Now, the $\Bar{h}$ is again so large that it quenches the superconducting state of the bottom Al layer. The superconductivity of the top Al layer is preserved, as can be concluded from the remaining peaks and the gap. Throughout the sweep of the $B$ field, the peaks shift almost linearly in $B$ as represented by the fit lines: (solid line) $eV_1 = \Delta_1 +\Delta_2+\Bar{h}= 580 \mu eV + 35\mu_B B$ and (dashed line) $eV_2 = \Delta_1 +\Delta_2-\Bar{h} = 290 \mu eV -35\mu_B B$. {From the difference between the two curves, it is possible to estimate $\Bar{h}(B)=145 \mu eV + 35\mu_B B$ and to extract $\Delta_1\simeq 240 \mu$eV. From the tunneling spectroscopy at 15~mT it is possible to evaluate $\Delta_2\simeq 200 \mu$eV}     
    (d) Tunneling spectroscopy at no applied magnetic field: before the first magnetization (black) and after magnetization (blue). Data extracted from (a) and (c) respectively, as indicated by the arrows.
    (e) \textbf{Magnetic field sweep of the tunneling spectroscopy at $T=1$ K.} At a higher temperature, the size of the superconducting gaps ($\Delta_{1,2}$) decreases and the spin-split peaks are only visible when the external magnetic field is opposed to the polarization of the EuS layer. The surface spins have enough energy at this temperature to be independent of the EuS  grains, explaining why the spin-splitting increases with increasing applied magnetic field.  
    A third, so called matching peak located between the superconducting gap and the spin-split peaks is likewise observed. With increasing negative $B$ field the peaks shift almost linearly in $B$ as represented by the fit lines: (brown solid line) $eV_1 =\Delta_1 +\Delta_2+\Bar{h}= 340 \mu eV - 172\mu_B B$; (blue dashed line) $eV_2 =\Delta_1 +\Delta_2-\Bar{h}= 250 \mu eV + 172\mu_B B$; (green dash-dotted line) $eV_3 =\Delta_1 -\Delta_2+\Bar{h}= 140 \mu eV -172\mu_B B$. {With a simple linear combinations of the three fit functions it is possible to estimate $\Bar{h}(B)=45 \mu eV - 172\mu_B B$, $\Delta_1\simeq 195 \mu$eV and $\Delta_2\simeq 100 \mu$eV with the assumption of a negligible dependence of $\Delta_{1,2}$ on $B$.} %
    (f) Tunneling spectroscopy at $B= -9.5$ mT, taken from panel (e). The spin-split peaks (brown and blue) and the matching peak (green) are visible (indicated by a circle of the respective color as corresponding to the fit lines in (e)). 
    } 
    \label{Fig._Inital-charac}
\end{figure*}

Characterization of the magnetic proximity effect was also necessary in the SuperTED project {and was quantified from the tunneling spectroscopy of Al/AlOx/Al/EuS samples of different thicknesses}. \cite{hijano_coexistence_2021}
Table~\ref{tab:table2} summarizes selected measured samples and the extracted values for the order parameter of the Al tunnel electrode ($\Delta_1$) and of the Al layer adjacent to the EuS ($\Delta_2$), as well as the 
induced exchange splitting $\Bar{h}$, and the ratio $\Bar{h}/\Delta_2$.
In Fig.~\ref{Fig._Inital-charac} we show an example of  tunneling spectroscopy of an Al/EuS bilayer in which the Al thickness was optimized to set $\bar{h}$ just below the maximal Chandrasekhar-Clogston limit. {It corresponds to sample S3-e in Table \ref{tab:table2} and it is partially reported in ref.~\cite{hijano_coexistence_2021}}.
The EuS/Al bilayer is in tunnel contact with an Al probe that allows us to quantify $\bar{h}$ from the amplitude of the splitting observed in the energy spectrum of the tunneling conductance. Notably, after the cooldown of the device and before the magnetization of the EuS layer a small splitting was already visible in the tunneling spectroscopy as shown in the black line of Fig.~\ref{Fig._Inital-charac}b. This indicates 
{the presence of}
magnetic domains with sizes larger than the superconducting coherence length $\xi_0$ in the EuS film as described in previous works~\cite{strambini_revealing_2017}. Domain sizes much below $\xi_0$ would lead to a vanishing average spin splitting. Upon the application of an external magnetic field ($B$) the magnetic domains start to align and the observed splitting increases as shown in Fig.~\ref{Fig._Inital-charac}a up to an in-plane field of 16 mT. Above this field the secondary peak at the higher energy suddenly disappears as a consequence of the S to N transition induced in the Al/EuS bilayer by the increased $\bar{h}(B)$.

The superparamagnetic component of the EuS is further confirmed by measuring the tunneling spectroscopy at lower magnetic fields as shown in Fig.~\ref{Fig._Inital-charac}c. The magnetic field response is hysteretic: By tracing $B$ back a N to S transition can be observed at $B<9$ mT with the re-appearance of the secondary peak. The splitting then decreases by lowering $B$ down to -12 mT. Below -12 mT the EuS layer reverses its polarization and $\bar{h}$ increases again above the Chandrasekhar-Clogston limit with the resulting quenching of superconductivity. By reversing the sweep this picture reverts as expected for a ferromagnetic system (data not shown). {Interestingly, by following the position of the two peaks in B we notice an almost linear decreasing of the splitting (as underlined by the black fit lines in Fig.~\ref{Fig._Inital-charac}c) consistent with a linear field dependence of $\Bar{h}= \Bar{h}(0)+ 35 \mu_B B$. $\Bar{h}(0)\simeq 145$ $\mu$eV represents the intrinsic exchange splitting induced by the ferromagnetism of the EuS while the linear part in $B$ is the superparamagnetic component ($35\gg 2$) and corresponds to the action of the external magnetic field on the polarization of the surface spins interacting with the Al layer and weakly coupled to the bulk EuS. In addition, the action of $B$ ($< 9$ mT) applied anti-parallel with respect to the EuS polarization results in a decreasing of $\Bar{h}(B)$.}

The tunneling spectroscopy of the junction at higher temperatures reveals an even richer and non-trivial dependence of the bilayer superconductivity on the external magnetic field. This dependence is especially relevant for extending the temperature range of the detectors. In particular, as shown in Fig.~\ref{Fig._Inital-charac}d, the tunneling spectroscopy measured at 1~K is characterized by a normal state for the EuS/Al bilayer for most of the magnetic fields explored including at zero magnetic field, while only when the polarization of $B$ is anti-parallel with respect to the polarization of the EuS layer ($-2$ mT $>B>-10$ mT), the bilayer is in the superconducting state. 
This behavior can be explained by an increase of $d_C(T,\bar{h})$ overcoming the physical thickness of the device at higher temperature as estimated in Fig.~\ref{fig:phase_diagram}(b) and the monotonic relation with $\bar{h}$.
In contrast to the low-temperature measurement, now the exchange splitting is enhanced by the external magnetic field $\Bar{h}(B)=\Bar{h}(0) -172 \mu_B B$ (notice the opposite slope in $B$) with a stronger superparamagnetism (larger slope) and a weaker intrinsic effect $\Bar{h}(0)\simeq 45$ $\mu$eV. This high-temperature scenario is likely related with a ferromagnetic to paramagnetic transition of the EuS surface spin pattern expected to have a transition around $\simeq 900$ mK~\cite{strambini_revealing_2017}, much below the transition temperature of the bulk EuS ($\simeq 17$ K). As a result of this transition, the surface spin alignment, determining the spin splitting in the Al, is only weakly coupled to the magnetization of the bulk EuS. %
Notably, at high temperature, the tunneling spectroscopy is enriched by a third peak representing the matching peak appearing at $eV \simeq \Delta_1 - \Delta_2 + \bar{h}$.
{The almost linear evolution of the three peaks in $B$ is reported in Fig.~\ref{Fig._Inital-charac}d and underlined by three fit lines.}

Finally, similarly to the low temperature spectroscopy, by repeating the measurement in the opposite field direction the features are reversed with respect to the magnetic field (data not shown) confirming the magnetic hysteresis of the device and the ferromagnetism of the EuS layer.

\begin{figure}[t]
    \centering
    \includegraphics[width=0.55\textwidth]{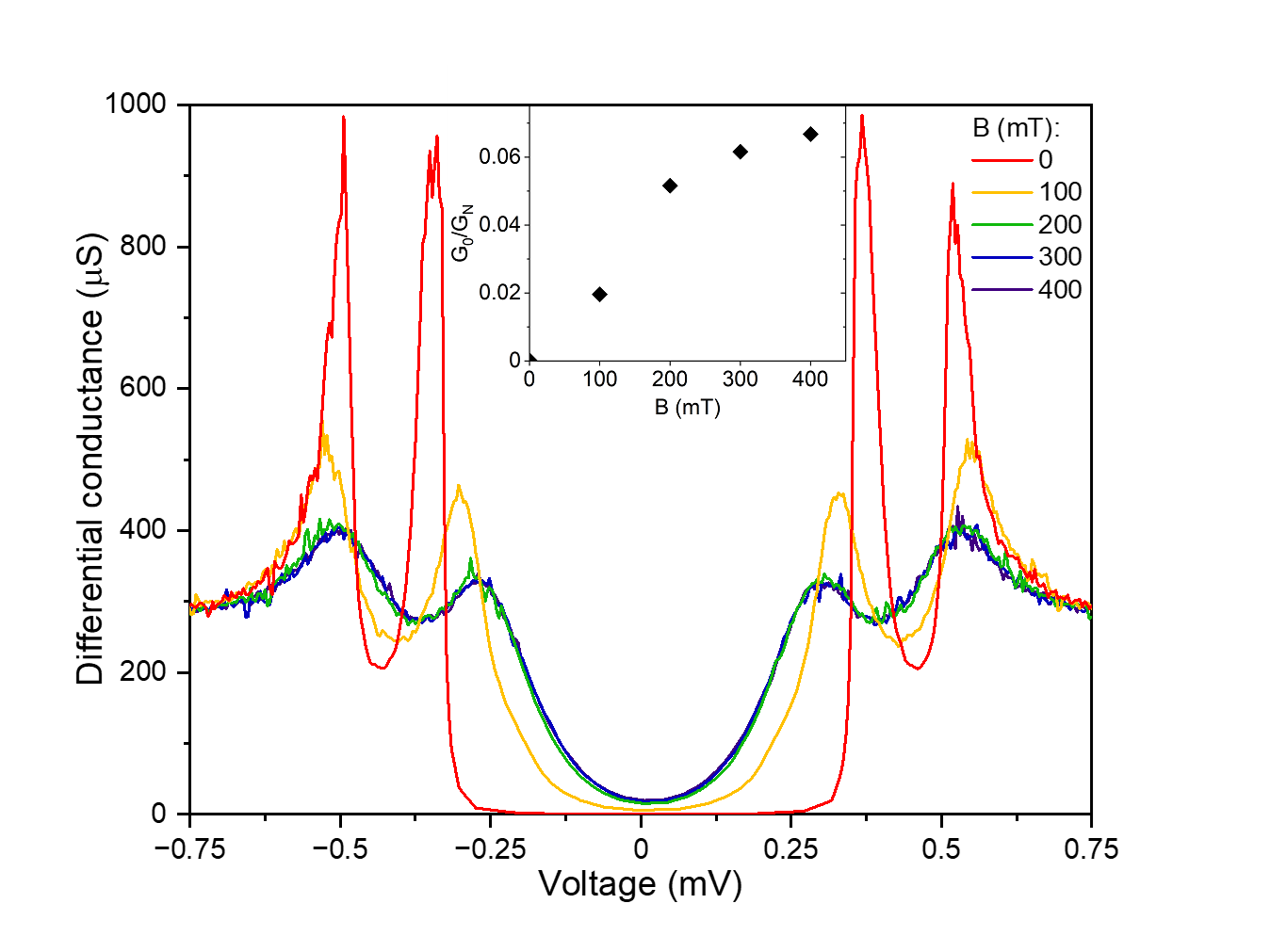}
\caption{\textbf{Differential conductance as a function of in-plane external magnetic field for sample S1-b.} Subgap states and delta vary with $B$. The inset shows the field dependence of the zero bias conductance $G_0(B)$ normalized respect to the normal state conductance $G_N$.}
\label{fig:B}
\end{figure}

Within the same batch of samples some devices (S1) showed a weaker $\Bar{h}$ so that superconductivity could survive at larger in-plane magnetic fields. In Fig.~\ref{fig:B} we show the evolution of the tunneling spectroscopy at large magnetic fields, up to 400 mT. Also for these devices a clear superparamagnetic effect is visible in the enhancement of the Zeeman splitting at large magnetic fields. 
Moreover, increasing the magnetic field also lowers the amplitude of the peaks and increases the zero bias conductivity, a phenomenon that can be described with an enhancement of spin-scattering and/or inelastic scattering events typically described by a generic Dynes parameter $\Gamma$~\cite{munzenberg2004superconductor,kolenda_observation_2016}. {The evolution of the zero-bias conductance ($G_0$) compared to the normal state conductance ($G_N$) is shown in the inset of Fig.~\ref{fig:B}.}
{Although the precise origin of the magnetic field dependence is not clear from this data, a possible explanation is the orbital depairing from the combination of the external field and the stray fields of the EuS.}

\begin{figure*}[t]
    \centering
    \includegraphics[width=0.8\textwidth]{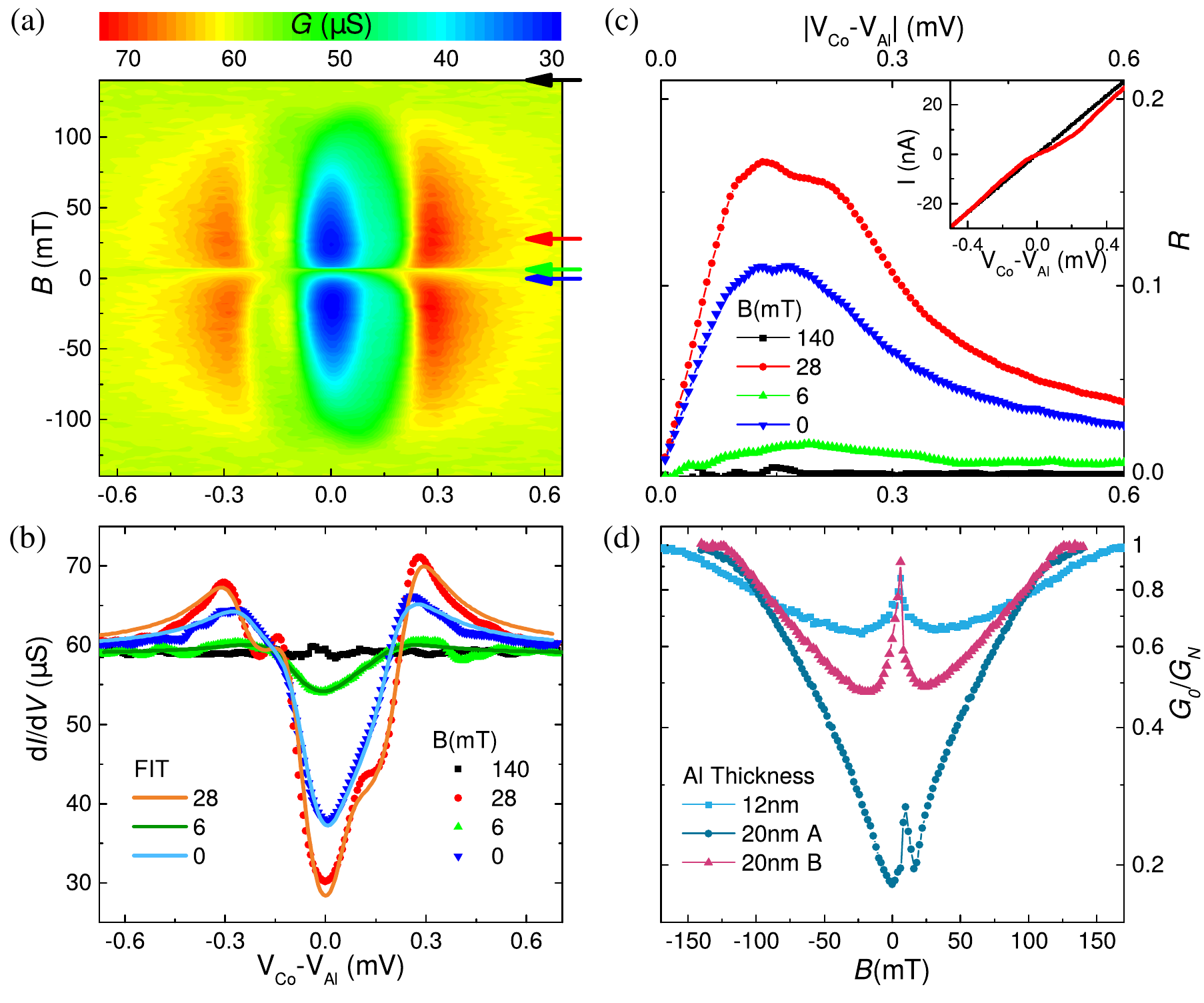}
    \caption{(a) and (b) Differential conductance measured for typical spin selective tunnel junction made of EuS/Al/AlOx/Co at different bias voltages $V$ and magnetic fields. The field was swept from -140 to +140 mT.
    (b) Traces selected from (a) showing: a persistency of the spin polarization and filtering also in the absence of magnetic field ($B=0$ blue dots); the maximal asymmetry at 28~mT (red dots); no asymmetry and spin filtering at the coercive field of EuS ($B=6$~mT green dots); and an almost flat tunneling spectroscopy corresponding to the quenching of superconductivity at large magnetic fields ($B=140$~mT green dots). 
    (c) Rectification coefficient $R=I_{\rm sym}/I_{\rm asym}$ extracted from the same IV characteristics used in panel (b). Notably, the $R(V)$ curves provide an alternative method to quantify the asymmetry of the junction and the degree of spin polarization as described in Sec.~\ref{subsec.S/FI transport}.
    (d) Field dependence of the Dynes parameter $\Gamma=G_0/G_N$ extracted from the tunneling spectroscopy of three similar samples of EuS/Al/AlOx/Co. Notably, the increase of $\Gamma$ at $B\simeq 6$~mT corresponds to the switching field of the EuS and Co layers, while at large fields ($|B|>100$mT) $\Gamma$ tends to unity as a consequence of the quenching of the superconducting layer. 
    }
    \label{Fig2DOS}
\end{figure*}

\begin{figure*}[t]
    \centering
    \includegraphics{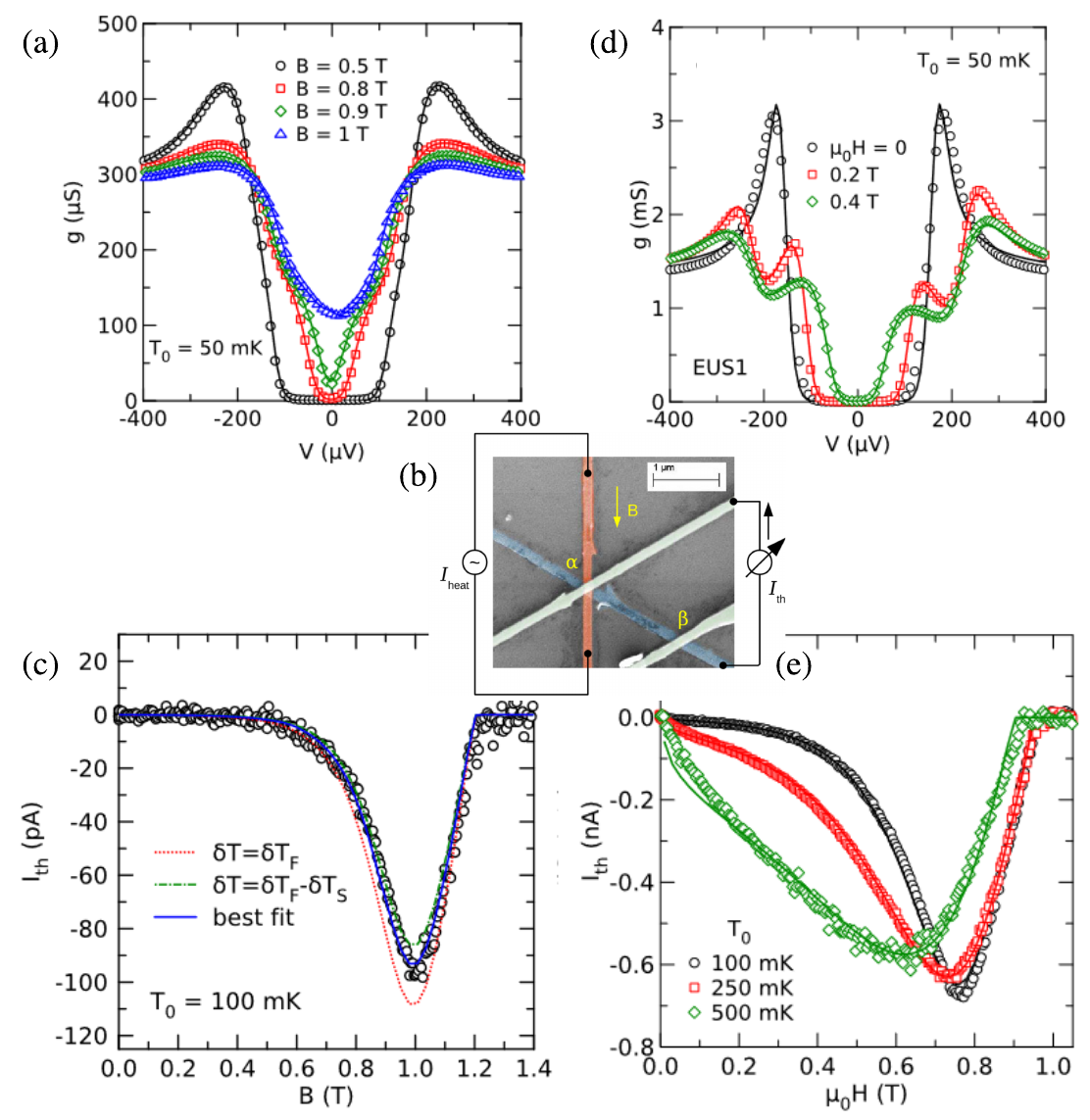}
    \caption{
    (a) Differential conductance of an Al/AlOx/Fe tunnel junction  as a function of voltage bias $V$ for different magnetic fields $B$.
    (b) False-color scanning electron microscopy image of a region of the Al/AlOx/Fe sample. The image displays the measurement configuration for thermoelectric measurements. 
    (c) Thermoelectric current $I_{th}$ as a function of magnetic field $B$ measured at a base temperature $T_0$ = 100 mK, together with theoretical fitting.  
    (d) Differential conductance of the tunnel junction EuS/Al/AlOx/Fe as a function of the applied bias $V$ for various applied fields.
    (e) Thermoelectric current $I_{th}$ measured for the sample in (d) as a function of the applied magnetic field $\mu_0 H$ at different base temperatures $T_0$.
    Data adapted from \cite{kolenda_observation_2016,kolenda_thermoelectric_2017}.
    }
    \label{Fig.Beckman}
\end{figure*}

\subsubsection{Spin selective tunnel junctions}

As shown in Sec.~\ref{sec1}, to obtain non-reciprocal transport from a FI/S based tunnel junction a spin selective probe is required. There are two primary methods to achieve this: (i) by using a FI as a spin selective tunnel barrier in S/FI/N~\cite{strambini2022superconducting, moodera_electron-spin_1988, hao_spin-filter_1990, moodera_variation_1993, santos_observation_2004} or S/FI/S \cite{hao_spin-filter_1990, santos_observation_2004, miao_controlling_2009, miao_magnetoresistance_2009, muller_exchange_2009} junctions using a ferromagnetic probe in 
FI/S/I/F~\cite{ strambini2022superconducting, moodera_frontiers_2010} and S/FI/F~\cite{thomas_evidence_2005} configurations. For both configurations, clear non-reciprocal transport has been observed with an $I-V$ characteristic of the junction similar to an ideal semiconducting pn diode for small voltage bias ($V < \Delta_0$), with corrections dependent on the non-ideal spin polarization ($ 1-\mathcal{P}$) as described in Eq.~\eqref{Eq:ThermRect}.
The non-reciprocity is also visible in the asymmetry of the tunneling spectroscopy as depicted in Fig.~\ref{Fig2DOS}a and b showing the differential conductance measured at different magnetic fields and bias voltages for an EuS/Al/AlOx/Co tunnel junction. 

From the amplitude of the asymmetric peaks it is possible to extract the spin polarization of the junction $\mathcal{P} \simeq 0.2$ at 28 mT and $\mathcal{P} \simeq 0.15$ at 0 mT, then showing a permanence of spin-polarization even at zero field.
Unlike tunnel barriers that employ non-ferromagnetic probes, the zero bias conductivity ($G_0$) is not much smaller than the normal-state conductance $G_N$, an effect similarly observed in prior experiments~\cite{li_observation_2013} and similar devices. Such subgap conductance depends also on the applied external magnetic field as shown in Fig.~\ref{Fig2DOS}d, showing $\Gamma(B)$ evaluated from the ratio between the normal-state and zero-bias conductances for three similar devices. 
Notably, for all the devices $\Gamma(B)$ shows a similar trend with a minimum at low magnetic fields parallel to the polarization of the two ferromagnetic layers while a local maximum appears at the coercive field up to the global maximum of unity above 100 mT and corresponding to the quenching of superconductivity.%

The origin of this subgap conductivity is still under debate. For example, stray fields from the Co layer as well as pin-holes in the AlOx barrier can impact the superconductivity of the Al layer. 
Independent of the microscopic origin of the subgap conductance, the net effect on the device performance is a reduction of non-reciprocal charge transport that, to the first order, is linear in $\Gamma$ as shown in Eq.~\ref{eq.DeltaI} while it is less detrimental (by a factor $hT^2/\Delta^3$) for thermoelectricity~\cite{ozaeta_predicted_2014} as shown in Eq.~\ref{eq.DeltaAlpha}.

\subsubsection{Thermoelectricity in spin selective tunnel junctions}

Thermoelectricity in spin selective tunnel junctions has been observed, for the first time, in Al/AlOx/Fe junctions~\cite{kolenda_observation_2016}. Here the Zeeman splitting of the Al layer was induced by an external in-plane magnetic field as shown in the tunneling spectroscopy measurement of Fig.~\ref{Fig.Beckman}a. The Fe contact was heated by a Joule current and thermoelectric current generated from such a thermal gradient was measured with a third Cu contact using a circuit scheme shown in Fig.~\ref{Fig.Beckman}b. With this configuration a thermocurrent up to 100 pA was observed at $B=1$ Tesla, 100 mK bath temperature and  $\simeq$100 mK thermal gradient (data reported in Fig.~\ref{Fig.Beckman}c extracted form Ref.~\cite{kolenda_observation_2016}). Notably, a sizable thermoelectric signal was visible only above 0.5 T, a field at which the tunneling spectroscopy shows an asymmetry, confirming the intimate relation between thermoelectricity and non-reciprocity.
From the measurement a Seebeck coefficient up to $\simeq 100$ $\mu V / K$ was estimated. The measured effects were in a very good agreement with the tunneling theory  of \cite{ozaeta_predicted_2014} discussed in Sec.~\ref{subsec.S/FI transport}. One year later, the same group performed a similar experiment on a EuS/Al/AlOx/Fe junction~\cite{kolenda_thermoelectric_2017}. Thanks to the presence of the additional EuS layer a larger thermometric signal (up to $0.7$ nA) was visible already at small magnetic fields due to the strong exchange interaction induced by the FI layer as shown in  Fig.~\ref{Fig.Beckman}e. However, thermoelectricity was not visible at zero magnetic field consistently with the weak ferromagnetic response of the EuS as observed also from the tunneling spectroscopy in Fig.~\ref{Fig.Beckman}d. This was likely due to the poor interface between EuS and Al resulting from the nanostructuring procedure that requires exposing EuS to air before the growth of the Al layer.

After adding an antenna to the setup of Fig.~\ref{Fig.Beckman}, the system could in principle be used as a thermoelectric detector. There the effect of transverse rectification in detection is likely negligible due to the fact that the lateral size of the junction is much smaller than the characteristic length estimated in \cite{strambini2022superconducting}. However, the operation of the detector is likely hampered by the necessity of applying a relatively large magnetic field to reveal the magnetic proximity effect.

\section{Applications}

{In this second part, we explore the applications of S/FI structures in radiation detection. In the first section, we describe how an SFTED can be used as a microcalorimeter for X-ray detection. The second section is devoted to THz detection, whereas in the third section, we discuss possible multiplexing read-out.  }

\subsection{X-ray detection\label{sec5}}

A microcalorimeter is a microscale thermal detector that thermalizes the energy of an absorbed particle, creating a temperature %
deviation which is then measured by a sensor. Modern microcalorimeters should possess high energy resolution, broad bandwidth, minimal dissipation, straightforward fabrication, and the capability for operation in a sensor-array with readout-multiplexing. The superconductor-ferromagnet heterojunction is an attractive sensing device for cryogenic microcalorimeter applications\cite{Chakraborty2018}. The pronounced thermoelectric effect\cite{ozaeta_predicted_2014,kolenda_observation_2016} in this device converts the absorbed energy into an electrical signal without requiring bias power, fundamentally reducing the heat dissipation and wiring complexity of the detector. Theoretical studies indicate that a superconductor-ferromagnet thermoelectric detector (SFTED) has the potential to be a swift cryogenic microcalorimeter. Its energy resolution could rival state-of-the-art ultrasensitive detectors, such as the transition-edge sensor (TES){\cite{Irwin1996,Ullom2015}}, kinetic inductance detector (KID)\cite{Grossman1991}, and the superconducting tunnel junction (STJ)\cite{Kurakado1982}.

This section reviews the linear microcalorimeter theory of the SFTED as presented in Refs.~\cite{Heikkila2018,Chakraborty2018,geng2020a}. We also address the time-domain analytical models and the pulse excitation solutions for the SFTED\cite{geng2022d}. Furthermore, design considerations and optimizations of the detector are discussed with numerical examples using practical parameters. 

\begin{figure*}[t]
    \centering
    \includegraphics[width=0.9\linewidth]{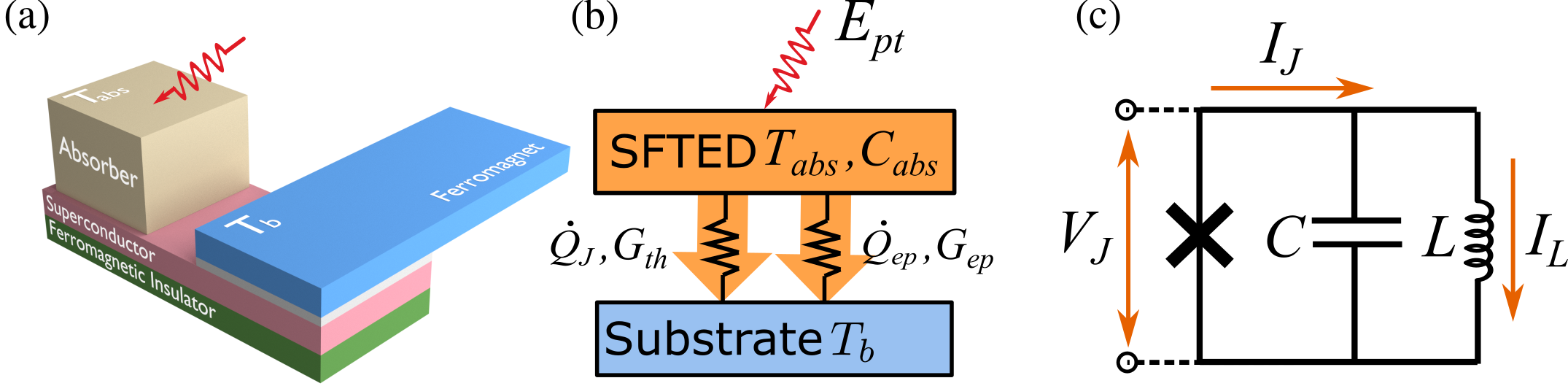}
    \caption{Schematic of a (a) superconductor-ferromagnetic thermoelectric detector (SFTED). (b) and (c) are the thermal model and the electrical circuit of the SFTED.}
    \label{fig:xray_schematic}
\end{figure*}

\subsubsection{Linear theory}
A simplified schematic of the SFTED is presented in Fig.~\ref{fig:xray_schematic}. Consider an X-ray photon with energy $E_{pt}$ impacting a superconducting absorber.
The thin superconducting electrode of the junction is assumed to be thermalized with the bulk absorber, thus sharing the same electron temperature $T_{\rm abs}$, which jumps above the phonon bath temperature $T_b$ after the absorption of an X-ray photon, and then decays exponentially due to the heat flow into the bath. Conversely, the normal electrode of the junction remains at the bath temperature $T_b$.

A finite temperature difference, $\Delta T=T_{\rm abs}-T_b$, induces a thermoelectric voltage $V_J$ across the junction, simultaneously creating a heat current $\dot{Q}_J(\Delta T,V_J)$ [see panel (b)] and a thermoelectric charge current $I_J(\Delta T,V_J)$ [see panel (c)], transporting energy and charge between the hot and the cold electrodes, until the detector is restored back to its quiescent state $\Delta T=0$.

The detector's thermal and electrical balance govern $\dot{Q}_J(\Delta T,V_J)$ and $I_J(\Delta T,V_J)$, and within the small-signal regime $\Delta T < T_b$, the balance can be expressed in a linearized form using the state variables $\Delta T$ and $V_J$ \cite{Heikkila2018,geng2020a} as:
\begin{equation}
    \begin{aligned}
        C_{\rm abs}\frac{\dd \Delta T}{\dd t}&=\dot{Q}_{in}-G_{th}^{\rm tot}\Delta T+\alpha V_J \\
        I_J & = \alpha\frac{\Delta T}{T_b} - \frac{1}{R_J}V_J,
    \end{aligned}
    \label{eq:xray_th_el_general_eqs}
\end{equation}
where $C_{\rm abs}$ is the heat capacity of the absorber\footnote{Due to the large volume difference between the absorber and the thin superconducting electrode ($<20$ nm) in practical devices, we assume that the absorber heat capacity dominates and omit the contribution of the electrode.}, $R_J\equiv\partial V_J/\partial I_J$ and $\alpha\equiv T_b\partial I_J/\partial \Delta T$ are the dynamic resistance and thermoelectric coefficient \cite{ozaeta_predicted_2014} of the junction at $V_J=0$ and $\Delta T=0$, respectively. In Fig.~\ref{fig:xray_schematic}(b)'s thermal model, $G_{th}^{\rm tot}=G_{th}+G_{ep}$ represents the total thermal conductance between the sensing elements — the absorber and superconducting electrode — and the bath, where $G_{th}\equiv\partial \dot{Q}_J/\partial \Delta T$ and $G_{ep}\equiv\partial \dot{Q}_{ep}/\partial \Delta T$ are the thermal conductances through the tunnel barrier of the junction and through the electron-phonon coupling in the superconductors, respectively. 

In the electrical model from Fig.~\ref{fig:xray_schematic}(c), the thermoelectric current $I_J$ at frequency $\omega$ is given by $I_J(\omega)=[i\omega C+1/(i\omega L)]V_J$, with $L$ and $C$ as the inductance and capacitance of the lumped circuit. As a result, Eqs.~\eqref{eq:xray_th_el_general_eqs} read as $\dot{Q}_{in}(\omega)=Y_{th}(\omega)\Delta T-\alpha V_J$ and $\alpha\Delta T/T_b=Y_{el}(\omega) V_J$ in frequency domain, where $Y_{th}=i\omega C_{\rm abs}+G_{th}^{\rm tot}$ and $Y_{el}=1/R_J+i\omega C+1/(i\omega L)$ are the total thermal and electrical admittances \cite{Heikkila2018}.

Thus, the voltage and current responsivity of the detector, considering capacitive and inductive loads, can be defined as:
\begin{equation}
    \lambda_V \equiv\frac{V_J}{\dot{Q}_{in}} = \frac{\alpha}{Y_{th}Y_{el}T_b-\alpha^2},\quad
    \lambda_I \equiv\frac{I_L}{\dot{Q}_{in}} = \frac{\lambda_V}{i\omega L},
\end{equation}
where $I_L$ is the current in the inductor.

The noise equivalent power (NEP) is a key metric to evaluate the performance of a detector, particularly important for bolometric applications. It is defined as the input radiation power in 1 Hz bandwidth required by the detector to generate a signal equal to its noise. For a SFTED, the total $\mathrm{NEP}_{\rm tot}^2=\mathrm{NEP}_{amp}^2 + \mathrm{NEP}_{\textrm{TED}}^2$, where $\mathrm{NEP}_{amp}$ is induced by the noise from the amplification chain that is used to readout the detector signal, and $\mathrm{NEP}_{\textrm{TED}}$ arises from the junction noise combined with the thermodynamic fluctuation noise (TFN) \cite{Heikkila2018}:
\begin{equation}          
    \mathrm{NEP}^2_{\textrm{TED}}=\frac{4k_BT_b^2G_{th}^{\rm tot}}{\mathrm{ZT}}\big[
    1+(1+\mathrm{ZT})\tau_{th}^2\omega^2\big],
    \label{eq:xray_NEPJ}
\end{equation}
where $\tau_{th}=C_{\rm abs}/G_{th}^{\rm tot}$ is the thermal time constant and ZT is the thermoelectric figure of merit reading as $\mathrm{ZT}(\omega)=\alpha\lambda_V(\omega)$. Here it is worth noting that a ZT value exceeding one enhances the NEP of SFTED over a standard bolometer at low frequency, and this comes about because of the direct negative electrothermal effect on the noise \cite{geng2020a}. 

Furthermore, it has been shown \cite{Heikkila2018} that the zero-frequency NEP$_\textrm{TED}$ can be optimized with respect to $R_T$ yielding a condition $G_{th}/G_{ep}=\sqrt{1+zT_i}$ where $zT_i=\alpha/(G_{th}T_b/R_J-\alpha^2)$ is the intrinsic figure of merit of the SFTED. As a result, the optimal detector intrinsic zero-frequency noise equivalent power reads as:
\begin{equation}
    \mathrm{NEP}^2_{\textrm{TED,opt}}=\frac{4k_BT_b^2G_{ep}}{\mathrm{zT_i}}\big(
    1+\sqrt{1+\mathrm{zT_i}}\big)^2, 
\end{equation}
which is an important result particularly in guiding the design of SFTED for the bolometric applications.

For calorimetric applications, the performance of the detector is widely benchmarked by the energy resolution $\Delta E$. In the small-signal limit it is \cite{Irwin2005}
\begin{equation}
    \Delta E=\bigg(
    \int_0^\infty\dd\omega\frac{2}{\pi\mathrm{NEP}^2}
    \bigg)^{-1/2},
    \label{eq:xray_dE}
\end{equation}
obtained by applying the optimal filter and assuming an infinite bandwidth amplifier. For SFTED, the intrinsic energy resolution $\Delta E_{\mathrm{TED}}$ (omitting the amplifier noise) can be simplified to \cite{Chakraborty2018}:
\begin{equation}
    (\Delta E_{\mathrm{TED}})^2=4k_BT_b^2C_{\rm abs}\frac{\sqrt{1+\mathrm{ZT}}}{\mathrm{ZT}}.
\end{equation}
As one can see, the intrinsic energy resolution of the SFTED can go below the level of a TFN-limited feedback-less calorimeter ($\Delta E_{\mathrm{TFN}}^2=4k_BT_b^2C_{\rm abs}$) when $\mathrm{ZT}\geq1.6$.

Furthermore, \cite{geng2022d} shows how Eqs.~\eqref{eq:xray_th_el_general_eqs} can be solved analytically in time-domain, providing a direct description of the potentially complex behavior of the signal pulse of SFTED and a guide for designing and optimizing the detector and readout components. For the sake of simplicity, here we only revisit a case that the detector signal is readout from the inductive load, i.e., by a cryogenic current amplifier based on Superconducting QUantum Interference Device (SQUID), and the capacitance in Fig.~\ref{fig:xray_schematic}(c) is omitted ($C\rightarrow0$). 

Because of the strong similarity between the thermoelectric effect and the electrothermal feedback of a transition edge sensor (TES) \cite{Irwin2005},  Eqs.~\eqref{eq:xray_th_el_general_eqs} can be rearranged to a first-order differential matrix format \cite{geng2022d} analogous to those of the TES \cite{Irwin2005}, reading as
\begin{equation}
	\frac{d}{dt}
	\begin{pmatrix}
		I_L \\
		\Delta T
	\end{pmatrix}
	= 
	\begin{pmatrix}
		-\tau_{el}^{-1} & \frac{\mathcal{L}_IG_{th}^{\rm tot}}{\alpha L} \\
		-\frac{\alpha R_J}{C_{\rm abs}} & -\tau_{I}^{-1} 
	\end{pmatrix}
	\begin{pmatrix}
		I_L \\
		\Delta T
	\end{pmatrix},
 \label{eq:xray_mtxeqs}
\end{equation}
where we have defined $\mathcal{L}_I=\alpha^2R_J/(G_{th}^{\rm tot}T_b)$ as an analog of the constant current-bias low-frequency loop gain of the TES, $\tau_{I}=\tau_{th}/(1-\mathcal{L}_I)$ as the constant current thermal time constant, and $\tau_{el}=L/R_J$ as the electrical time constant. It is worth mentioning that, despite the similarity, the main differences between Eqs.~\eqref{eq:xray_mtxeqs} and those of TES \cite{Irwin2005} are in the definition of $\mathcal{L}_I$, and that the thermoelectric $\alpha$ appears in a dual role both within $\mathcal{L}_I$ and as the analog of the DC current of the TES. In particular, the thermoelectric factor $\alpha$ has a unit of current, making $\mathcal{L}_I$ correctly dimensionless. One should also note that it is totally different from the dimensionless logarithmic temperature sensitivity of the resistance for the TES, which is also typically denoted by $\alpha$. 

Equations \eqref{eq:xray_mtxeqs} have analytical solutions \cite{geng2022d}, which yield two time constants, denoted as $\tau_{\pm}^{-1}=(\tau_{el}^{-1}+\tau_{I}^{-1}\pm\sqrt{\Delta})/2$, where $\Delta=(\tau_{el}^{-1}-\tau_{I}^{-1})^2-4\mathcal{L}_I(\tau_{el}\tau_{th})^{-1}$. These time constants signify the rise ($\tau_+$) and decay ($\tau_-$) time of the SFTED current pulse in response to a delta-impulse absorption event. The general stability condition of a microcalorimeter detector requires that the signal pulse decays back to the quiescent state either with oscillation (underdampled case $\Delta <0$) or without (critically damped $\Delta=0$ or overdampled $\Delta>0$), requiring the real part of $\tau_\pm$ to be positive. An unbiased SFTED calorimeter is always stable because the loop gain satisfies $\mathcal{L}_I<1$ \cite{Heikkila2018,geng2022d}. However, an oscillatory signal extends the recovery time and significantly slows down the detector, therefore a more restrictive exponential pulse decay condition $\Delta\geq0$ is desired for the SFTED.

\subsubsection{Numerical results}

\begin{figure*}[t]
    \centering
    \includegraphics[width=0.8\linewidth]{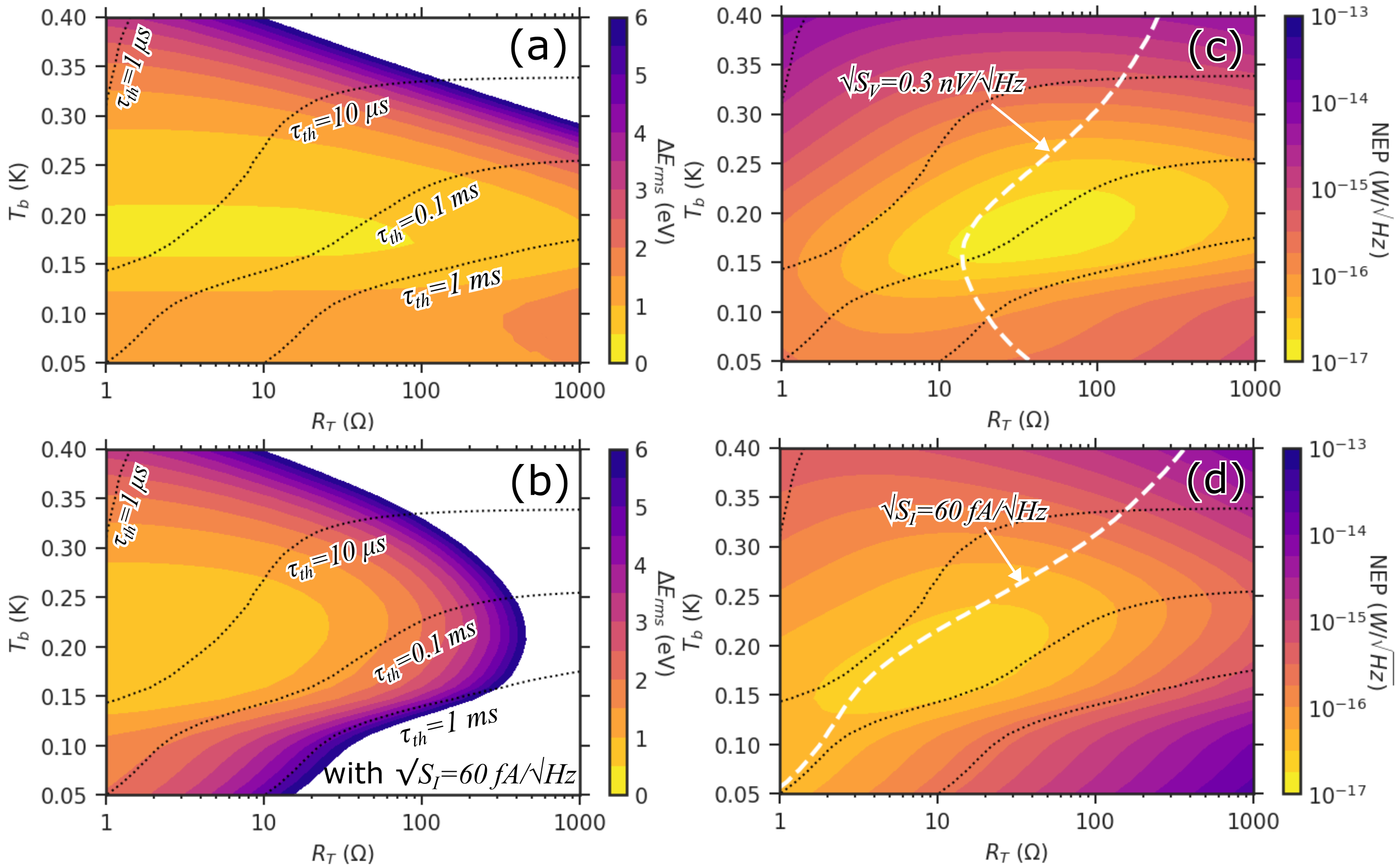}
    \caption{Color scaled detector energy resolution $\Delta E$ as a function of bath temperature $T_b$ and tunnel resistance $R_T$ (a) without the read-out noise and (b) with a current read-out that has current noise  
    $\sqrt{S_I}=60$ fA/$\sqrt{Hz}$. The black dotted lines denote four thermal time constant contour lines. The total noise equivalent power NEP$_{\rm tot}$ at 10 kHz of the detector read out by (c) voltage pre-amplifier with voltage noise spectral density $\sqrt{S_V}=0.3$ nV/$\sqrt{Hz}$ and (d) current pre-amplifier with current noise $\sqrt{S_I}=60$ fA/$\sqrt{Hz}$. The white dashed lines denote where the detectors intrinsic NEP$_{\mathrm{TED}}$ equals the NEP$_{\rm amp}$ contributed from the respective read-out, i.e. $\mathrm{NEP}_{\mathrm{TED}}=\sqrt{S_V}/|\lambda_V|$ is satisfied in panel (c) and $\mathrm{NEP}_{\mathrm{TED}}=\sqrt{S_I}/|\lambda_I|$ is satisfied in panel (d).}
    \label{fig:xray_dEvsRandT}
\end{figure*}

\begin{figure*}[t]
    \centering
    \includegraphics[width=0.7\linewidth]{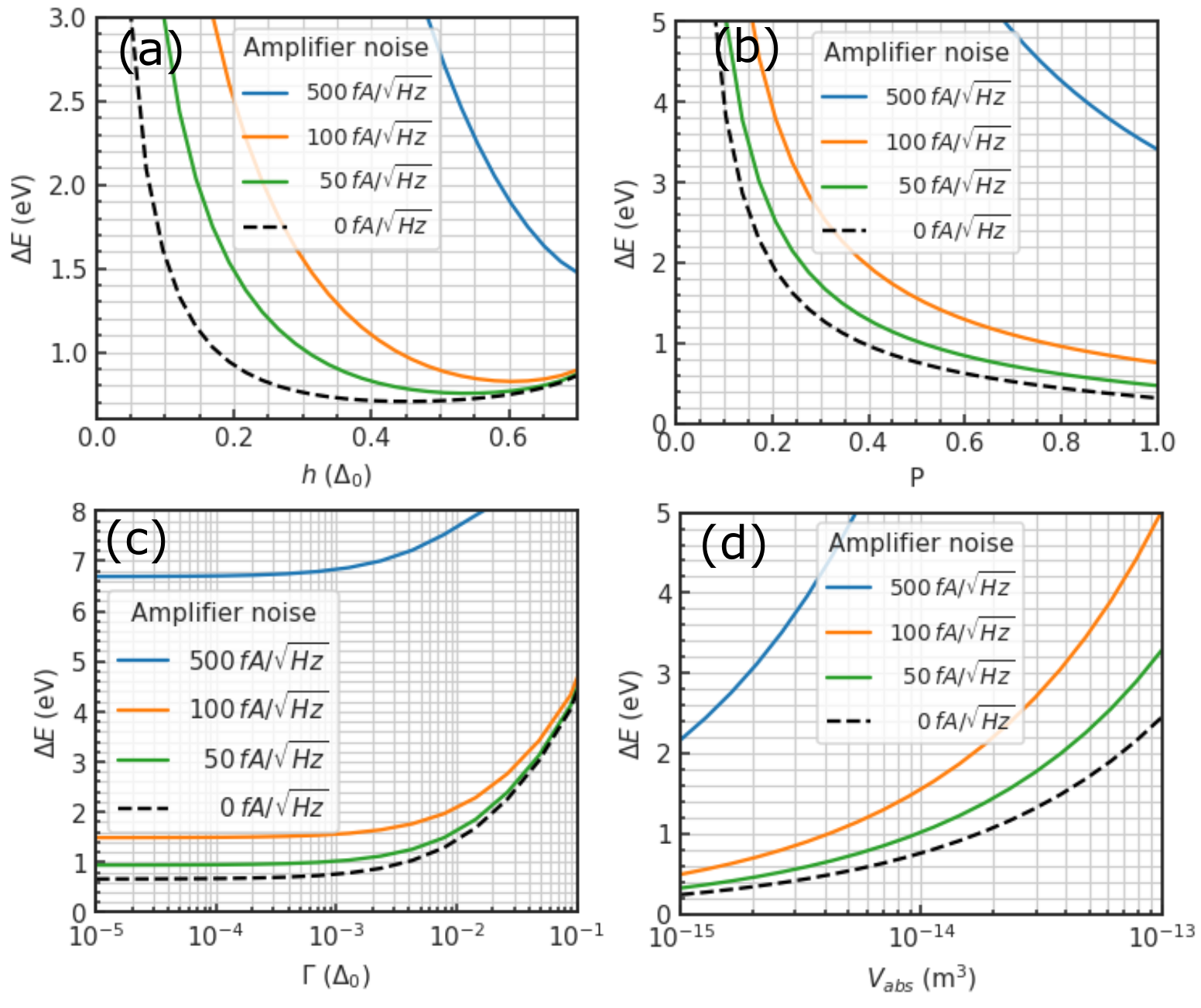}
    \caption{Total energy resolution $\Delta E$ (including amplifier noise) plotted as a function of (a) exchange field $h$, (b) polarization $P$, (c) broadening parameter $\Gamma$ and (d) absorber volume $V_{\rm abs}$. The three colored lines demonstrate $\Delta E$ with three different current read-out noise levels, and the black dashed line demonstrates the intrinsic resolution. Panel (d) is computed with materials parameters for an Sn absorber.}
    \label{fig:xray_dEvsParams}
\end{figure*}

\begin{figure*}[t]
    \centering
    \includegraphics[width=0.8\linewidth]{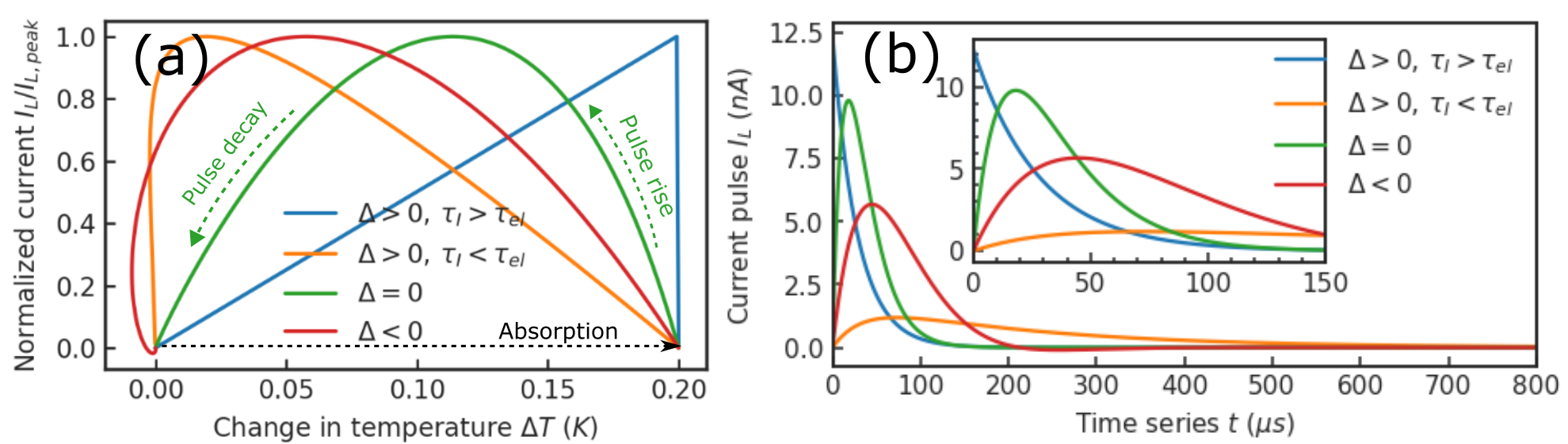}
    \caption{(a) Current-temperature pulse cycles under different operational conditions, denoted by the color of the lines. (b) Signal current pulse under different operational conditions corresponding to panel (a).}
    \label{fig:xray_pulseshape}
\end{figure*}

It is a non-trivial task to achieve the best performance of a detector. 
This is also the case when designing a SFTED. In particular, the non-ideality of the fabricated junction and the noise of the readout circuit and the amplifier will degrade the performance of a SFTED compared to the optimal operation regime. These topics have been analyzed and discussed in previous studies \cite{Heikkila2018,Chakraborty2018,geng2020a,geng2022d}. {Here we revisit the key conclusions with numerical examples that incorporate practical detector parameters. To achieve this, we employed numerical computations based on Eq.\eqref{eq:xray_dE} to determine energy resolution, as demonstrated in Fig.\ref{fig:xray_dEvsRandT} and Fig.\ref{fig:xray_dEvsParams}. Furthermore, we examine the transient pulses by numerically solving the eigenfunction Eq.\eqref{eq:xray_mtxeqs}, as illustrated in Fig.\ref{fig:xray_pulseshape}.
}

The most important parameters for designing a SFTED include: the tunneling resistance $R_T$, the exchange field $h$, the polarization $P$ and the broadening parameter $\Gamma$ of the tunnel junction, the volume $V_{\rm abs}$ and heat capacitance $C_{\rm abs}$ of the absorber, the noise power spectra density (PSD) of the read-out amplifier, and the temperature $T_b$ of the bath.

Among these parameters, the tunneling resistance $R_T$ and the bath temperature $T_b$ are of paramount importance to the optimization of the detector, as they are technically relatively easy to adjust. $T_b$ can be controlled by the cryogenic apparatus on-the-fly, whereas $R_T$ can be tuned in detector fabrication by modifying the area of the junction and/or the oxidation time of the tunnel barrier \cite{Gonzalez_2023}. In Fig.~\ref{fig:xray_dEvsRandT}, the theoretical energy resolution and the thermal time constant of a SFTED are demonstrated as functions of $R_T$ and $T_b$, when a set of realistic parameters, $h=0.3\Delta_0$, $P=0.5$, $\Gamma=10^{-3}\Delta_0$, $C=10$ nF, $L=2.6$ $\mu$H and $V_{\rm abs}=10^4$ $\mu$m$^3$ are selected to represent a real device. 

The intrinsic energy resolution of the SFTED $\Delta E_{\textrm{TED}}$ is presented in panel (a) together with four contour dotted-lines denoting different thermal time constants as functions of $R_T$ and $T_b$. Optimal SFTED performance is achieved around a temperature of $0.2$ K with a tunneling resistance under $100$ $\Omega$. Increasing or decreasing the temperature leads to a reduced figure of merit $ZT$ (see Fig.~2 panel (b) of Ref.~\cite{geng2020a}), whereas larger $R_T$ degrades $\Delta E$ by limiting the bandwidth of the detector.  

The electrical signal from a cryogenic microcalorimeter is typically too small to be directly read out by room temperature electronics. Consequently, a low-temperature pre-amplification of either the voltage or the current signal is required. Typically, voltage read-out works better for high-impedance sensors, whereas low-impedance sensors prefer current read-out. For a SFTED, it is not immediately obvious which type of read-out works better as $R_T$ can be adjusted via fabrication, and the total energy resolution $\Delta E$ depends on the noise of the amplifier. 

The total noise equivalent power NEP$_{\rm tot}$ of the detector at 10 kHz as functions of $T_b$ and $R_T$ are plotted in panel (c) when  using a state-of-the-art cryoHEMT \cite{jin2016} as voltage read-out ($\sqrt{S_V}=0.3$ nV/$\sqrt{{\rm Hz}}$, assuming $L\rightarrow\infty$) and in panel (d) for using a transformer coupled SQUID \cite{drung2007,wolf2017} as current read-out ($\sqrt{S_I}=60$ fA/$\sqrt{\rm Hz}$, assuming $C\rightarrow\infty$). In both panels, the white dashed lines signify the condition $\textrm{NEP}_{\textrm{TED}}=\textrm{NEP}_{\rm amp}$ such that the amplifier noise dominates the NEP$_{\rm tot}$ on the right (left) side of the dashed line when using a voltage (current) read-out.

When compared to the intrinsic energy resolution presented in panel (a), one can find that a SFTED with $R_T$ less than $100\textrm{ $\Omega$}$ with $T_b$ between $0.2\sim0.25$ K provides the best performance [$\Delta E_{\textrm{TED}}<2$ eV with $>10$ kHz bandwidth, see panel (a)] for a voltage read-out. Although with higher $R_T$ the noise is dominated by the detector's intrinsic noise and a smaller NEP$_{\rm tot}$ is obtained [see panel (c)], it leads to an undesired overall performance because of either a degraded intrinsic energy resolution [see panel (a)] or a large thermal time constant. As for a current read-out, it can be seen that the read-out noise matches to that from a SFTED with $R_T$ around $10\sim40\textrm{ $\Omega$}$ at $T_b$ between $0.2\sim0.25$ with a performance similar to the case of the voltage read-out. As an example, panel (b) shows the total energy resolution $\Delta E$ of a SFTED using a current read-out with noise of $\sqrt{S_I}=60\textrm{ fA/$\sqrt{Hz}$}$. One finds that the energy resolution  degrades drastically in the parameter region with higher $R_T$, the effect being stronger for low $T_b$. Clearly, optimizing between $T_b$ and $R_T$ is critical. 

In general, current read-out offers a larger bandwidth, has lower dissipation \cite{yvon2002} and good multiplexing schemes \cite{Ullom2015}. Hence it is preferable to be used with SFTED. As a result, to demonstrate the influence of other detector parameters in the following section, we consider a current read-out and set the detector to have $R_T=40 \textrm{ $\Omega$}$ at a temperature of $T=0.23$ K, based on the above discussion.

The total energy resolution, including the amplifier noise, is demonstrated as a function of the exchange field $h$ and the polarization $P$ in  Fig.~\ref{fig:xray_dEvsParams} panels (a) and (b), respectively,   with different read-out noise levels denoted by different colors. As discussed in Sec.~\ref{subsec:SFI}, $h$ and $P$ characterize the ferromagnetic-superconductor junction. They are influenced by both the ferromagnetic insulator and the ferromagnetic electrode materials, as well as by junction fabrication conditions.
Optimal resolution appears to be achieved when $h = 0.4-0.5$ and $P$ is maximized, which aligns with findings from Refs.~\cite{Heikkila2018,Chakraborty2018}.%

The broadening parameter $\Gamma$ of the tunnel junction, describing the broadening of the superconducting DOS and the excess sub-gap current, also influences the energy resolution, as discussed in Ref.~\cite{geng2020a}. As shown in Fig.~\ref{fig:xray_dEvsParams} (c), a higher %
$\Gamma$ leads to a degraded energy resolution. To achieve, for example, $\Delta E<2$ eV with an amplifier with $\sqrt{S_I}=100$ fA/$\sqrt{Hz}$, $\Gamma$ needs to be smaller than $10^{-2}\Delta_0$.

The absorber of a microcalorimeter absorbs the incident photon, and transfers and relaxes its energy to the electron and phonon systems within $10^{-9}$ s, the precise time depending on the absorber material  and size \cite{Zehnder1995,nussbaumer2000,ladstadter2004}. %
In Fig.~\ref{fig:xray_dEvsParams}(d) we demonstrate the dependence of $\Delta E$ on the volume $V_{\rm abs}$ of a superconducting Sn absorber. %
This takes advantage of the weak electron-phonon coupling in a superconductor, reducing heat loss to the bulk phonons \cite{Kaplan1976}. In contrast, a normal-metal absorber would often necessitate a more intricate micro-machined membrane or beam support structure to confine the absorbed energy in the phonon system \cite{Ullom2015}. Superconducting absorbers have been used in microcalorimetry applications \cite{perinati2004,irimatsugawa2015}, in particular together with superconducting tunnel junction sensors \cite{gaidis1996,Segall2004}. Generally speaking, the volume $V_{\rm abs}$ of the absorber should typically be determined by the desired X-ray energy range of the application such that the temperature change %
of the sensor induced by the absorption would stay in the linear response range of the detector, i.e., $E_{\textrm{x-ray}}/C_{\rm abs}<0.1T$  where $E_{\textrm{x-ray}}$ is the energy of the X-ray photon. For example, a Sn absorber with a volume $\sim10^{-14}$ m$^3$ has $C_{\rm abs}\sim10^{14}$ J/K \cite{oneal1965}, and thus can be used for the detection of an X-ray photon with keV energy. Moreover, it is also worth mentioning that the SFTED likely has weaker constraints on $V_{\rm abs}$ than a TES, as %
the transition to the non-linear regime is not as abrupt. 

We now illustrate the transient evolution of the detector signal in Fig.~\ref{fig:xray_pulseshape}. Panel (a) displays the pulse cycles in the signal current-temperature space ($I_L$-$\Delta T_J$) under four distinct conditions, whereas the corresponding current pulse signals in time domain are plotted in panel (b). With these examples,
SFTED has a desirable response when $\Delta>0$ (overdamping) and $\tau_{I}>\tau_{el}$, triggering faster pulses (blue curves). In contrast, two undesired conditions are also demonstrated with the orange ($\Delta>0$ but $\tau_{I}<\tau_{el}$) and red ($\Delta<0$) curves, both leading to slow response and recovery due to the reverse self-biasing \cite{geng2022d}, which should be avoided. On the other hand, the green curve represents the condition $\Delta=0$ wherein double roots emerge from Eq.~\eqref{eq:xray_mtxeqs}. Such a condition is often considered as an optimized compromise between the energy resolution and the slew rate requirement (finite detector rise-time) of the read-out electronics \cite{Irwin2005}.

\subsubsection{X-ray detector design}

\begin{figure}[t]
    \centering
    \includegraphics[width=0.9\columnwidth]{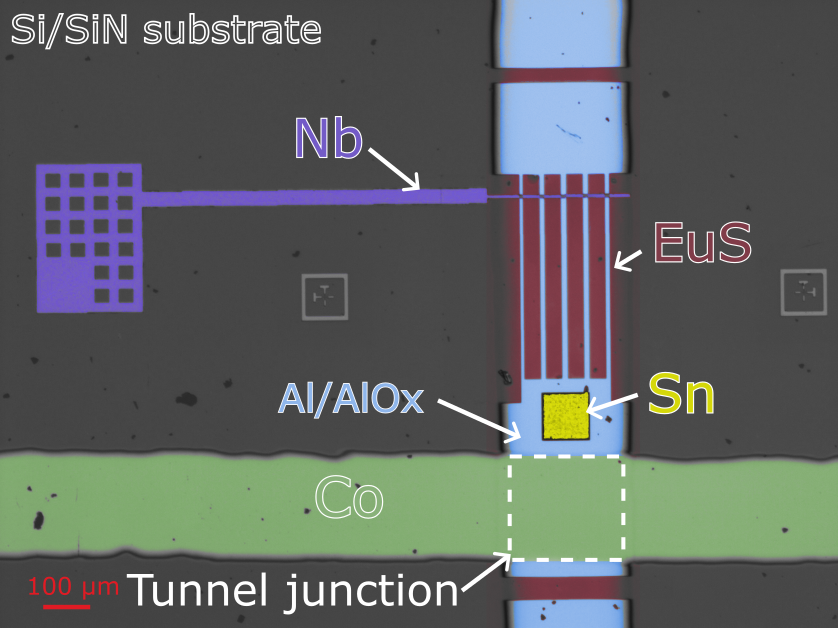}
    \caption{Fake-color optical micrograph of a SFTED calorimeter, in which the material compositions are illustrated with different colors. A superconductor-ferromagnet heterojunction area is marked with the white dashed line, consisting of layers of Si/SiN/EuS/Al/AlO$_x$/Co from the bottom to the top. A Sn absorber is placed to the side of the junction, contacting directly the Al lead with AlO$_x$ layer removed via chemical etching.}
    \label{fig:xray_detector}
\end{figure}

In the operation of an X-ray microcalorimeter, detecting a photon typically involves the detector to first absorb the energy of the incident photon, converting it to thermal excitations, and then to confine the excitations temporarily, creating a finite temperature excursion pulse. At the same time,  these excitations need to be transduced to a measurable electrical signal. For a practical device, these functions are carried out by the absorber, the thermal isolation and the sensing structures, respectively \cite{Ullom2015}.  Figure \ref{fig:xray_detector} shows an example of a real SFTED microcalorimeter, designed for X-ray detection. 

In this device, the sensing structure of the detector is a superconductor-ferromagnet heterojunction, consisisting of layers of EuS/Al/AlO$_x$/Co (from bottom to top), on the top of a nitridized Si substrate. %
Such a tunnel junction is fabricated by depositing a $200\textrm{ $\mu$m}$-wide ferromagnetic (Co) electrode on the top of and orthogonal to a $200\textrm{ $\mu$m}$-wide superconducting (Al) electrode with a thin plasma oxidized AlO$_x$ layer in between serving as the insulating tunnel barrier. The junction parameters are optimized for the detection of X-rays based on the discussion presented in previous sections.

The absorber structure is a $350$ nm thick superconducting Sn square (with an area of $100\textrm{ $\mu$m}\times100\textrm{ $\mu$m}$) providing 90\% and 13\% absorption efficiencies for 1 keV and 6 keV photons, respectively \cite{henkeb1993}. It contacts directly the superconducting Al electrode (with the insulating barrier AlO$_x$ removed chemically) and is placed laterally next to the junction area. Such a superconducting absorber configuration has been adopted before by superconducting tunnel junction (STJ) \cite{gaidis1996} and TES detectors \cite{zink2006}, and has been extensively studied both numerically and experimentally \cite{Segall2004}. 

Lastly, to confine the quasiparticles in the Sn absorber and Al electrode (the thermal excitations in an SFTED) and to minimize their excess escape, the Al electrode is patterned into an island with a lateral size of about the size of the absorber+junction, separated from the remaining leads, as illustrated in Fig.~\ref{fig:xray_detector}. In addition, a superconducting Nb lead is used to contact to the Al island via a set of narrow fingers. Due to the larger superconducting energy gap of Nb (typically $>1$ meV) contrasting to that of Al (typically $\sim0.2$ meV), such a lead provides an electrical connection, while effectively trapping the quasiparticles inside the Al island \cite{Segall2004}. On the other hand, because the temperature difference $\Delta T$ between the superconducting and ferromagnetic electrodes produces the electrical signal \cite{Heikkila2018} in a SFTED, the Co electrode is kept with a large area of $200\textrm{ $\mu$m}\times4\textrm{ mm}$, helping it to thermalize to the bath. 

\subsection{THz detection \label{sec6}}

The SFTED detector can be applied to measure electromagnetic radiation in a wide range of the spectrum. Besides X-rays, we consider here the measurement in the THz range. We define in this context the THz range as the range of frequencies comprised between 0.1 and 3~THz. In terms of wavelengths this is equivalent to a range of 0.1 to 3~mm. This region of the electromagnetic spectrum is particularly interesting for present and future astrophysics applications. Specifically, this includes observations of galaxy clusters, primordial galaxies, cold interstellar dust, and star-forming regions in our galaxy as well as neighboring galaxies \cite{Adam2018}. High-precision observations will in particular be needed in the coming years to be combined and complement with the exceptional JWST (James Webb Space Telescope) infrared data. 

To enhance the sensitivity of the SFTED junctions to incoming THz radiation, we have investigated several options. Among them, in decreasing order of fabrication complexity: 
\begin{itemize}
    \item Implementing a higher gap superconducting (e.g. Nb) planar antenna, feeding the SFTED sensor. From the fabrication point-of-view, this would have required adding a further metal layer (S2), and realising high-quality metal-on-metal contacts. On top of that, in order to properly match the typical antenna impedance with the SFTED junction, a lithography and etching of the SFTED structure would have been required.
    \item Patterning the SFTED junction into an impedance-matched absorber. The focusing of the THz radiation is achieved by adopting a micro-lens on the back of the silicon substrate. We stress that the dielectric substrate is transparent to THz radiation, allowing back-illumination. This approach is compatible with the existing layers, but would have required an additional, delicate lithography and etching of the SFTED junction structure.
    \item Based on the previously available shadow masks, post-patterning a non-optimised absorber in the existing S wire leading to the SFTED junction. This only requires a lithography on the superconducting (Aluminium) wire, away from the SFTED junction.
\end{itemize}

After discussions, we concluded that the last option was the safest for an initial proof-of-concept. For this purpose, a photosensitive resist, type S1805, is spun onto the chip and then baked. With UV lithography the pattern is subsequently realized. For both the development and the etching of the Aluminium, the developer solution MF319 is employed.

\begin{figure}[t]
\begin{center}
\resizebox{8cm}{!}{\includegraphics{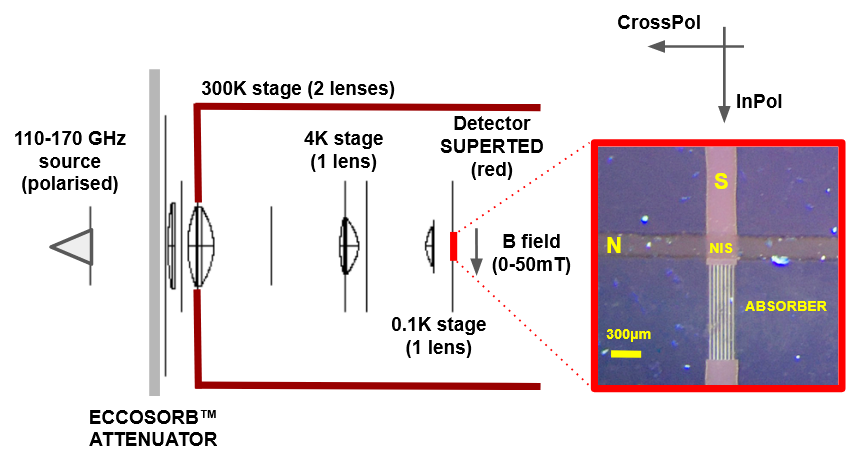}}
\caption{THz detector testing setup (top view) and picture of the SFTED junction. The cryostat optics is described in \cite{Monfardini2011}.}
\label{THz_setup}
\end{center}
\end{figure}

Figure \ref{THz_setup} shows a picture of the SFTED junction environment that has been fabricated and tested, together with a schematic of the optical cryogenics measurement setup. 

\begin{figure}[t]
\begin{center}
\includegraphics[width=0.7\columnwidth]{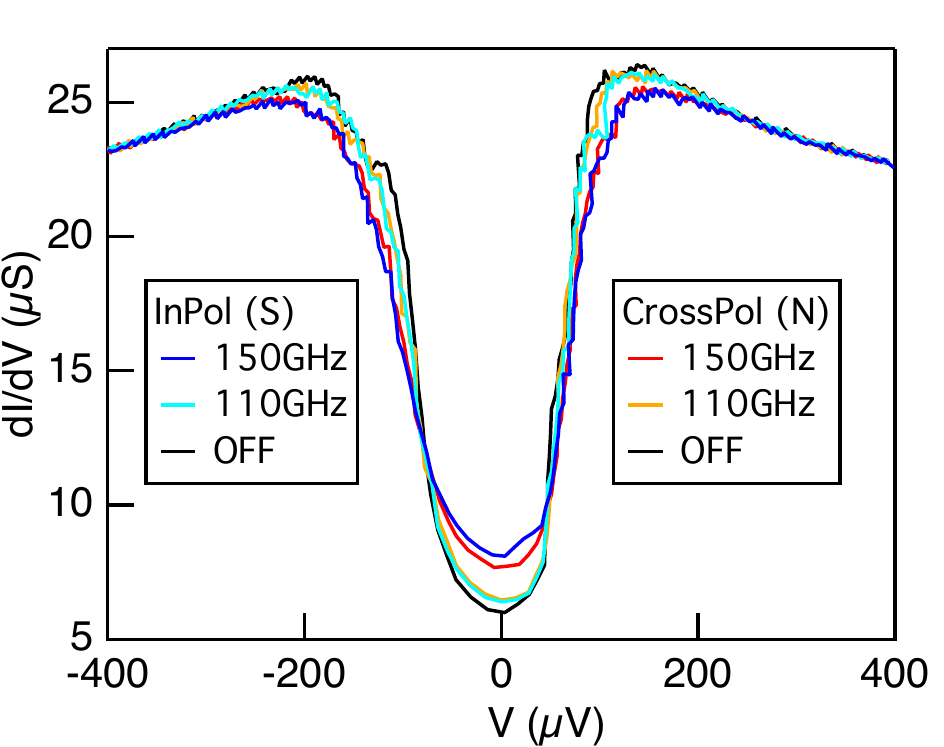}
\caption{Conductance-voltage characteristics of the SFTED junction shown in Fig.~\ref{THz_setup} under different illumination conditions. In particular, varying the direction of the polarisation and the frequency of the millimetre-wave source. $H~=~20$ mT along the direction of the superconductor (S) and $T$~=~120~mK. }
\label{THz_IV}
\end{center}
\end{figure}

\begin{figure}[t]
\begin{center}
\includegraphics[width=0.7\columnwidth]{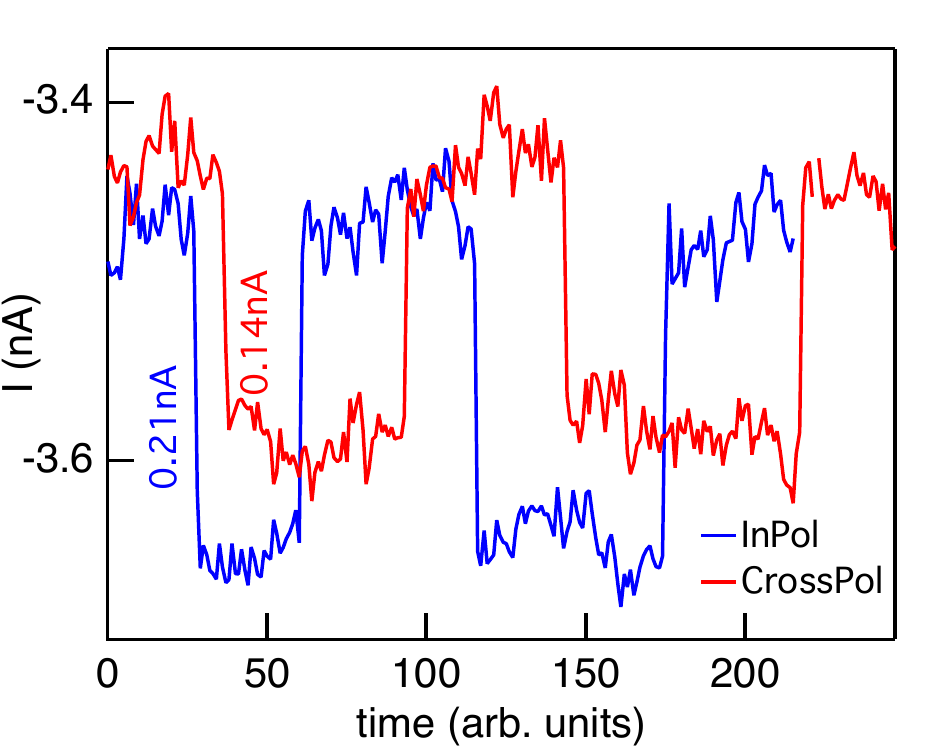}
\caption{Zero bias current flowing into the N electrode as a function of the 150~GHz illumination conditions (polarisation, and ON/OFF). $H$~=~20 mT along the direction of the superconductor (S) and $T$~=~120~mK.}%
\label{THz_TT}
\end{center}
\end{figure}

The preliminary results of the testing under illumination at 150~GHz and 110~GHz along the InPol (parallel to the S and the absorber wires) and CrossPol (parallel to the N wire) electric field polarisation directions are shown in Figs.~\ref{THz_IV} and \ref{THz_TT}.

In our initial measurement, we traced the SFTED junction's conductance-voltage characteristic under various illumination conditions. This measurement has been taken under an applied magnetic field of 20 mT along the direction of the superconductor (S). The millimeter-wave radiation distinctly impacts the curves, especially the residual conductance within the superconducting gap.

In a subsequent experiment, we left the detector unbiased and observed the current flowing into the N electrode. This represents the standard way for a SuperTED detector. We plot in Fig.~\ref{THz_TT} the time-traces of the current, expressed in nA, while switching ON/OFF the millimetre-wave source, and switching its polarisation between InPol and CrossPol. The ON/OFF steps, despite the presence of an offset current, represent the first hint of detection of 150~GHz photons from an un-biased SFTED junction, possibly related to the thermoelectric effect. Although the signal has a low S/N ratio, it appears to be roughly 50 \% stronger in the InPol (absorber, S) direction than in the CrossPol (N) direction. This is in nice agreement with the 3D electromagnetic simulations that we have run on the structure shown in Fig.~\ref{THz_setup}. These simulations predict the expected absorption of the metallic structure in the range 100-300~GHz.

Based on this first preliminary detection, the SuperTED detector, after decisive optimisations, might have the potential to compete with the existing cryogenic sensors operating at THz frequencies, i.e. Kinetic Inductance Detectors (KID), Transition Edge Sensors (TES), high-impedance bolometers and others. To our knowledge the preliminary data shown here is the first demonstration of  biasless superconductor-based radiation sensing. However, further work is still needed for its analysis and optimization. %

\subsection{Multiplexing read-out of self-powered detectors\label{sec7}}

The thermoelectric detectors TED are expected to generate a current upon absorbing electromagnetic radiation. In view of addressing hundreds of detectors at cryogenic temperature, we developed an idea for a possible highly multiplexed readout. Such a readout would solve both the complexity and the thermal load created by the hundreds of wires needed to address individual detectors. The readout is based on planar superconducting resonators biased by the DC-like current generated by the TEDs.

\subsubsection{Principle and design} 
A superconducting resonator can be modelled as an LC-circuit in which part of the inductance is coming from the kinetic inductance $L_k$ of the superfluid: the entire superfluid condensate opposes changes in the current direction. The resonator resonates at a frequency $f=1/(2\pi\sqrt{LC})$, where  $C$ is the capacitance and $L=L_K+L_G$ is the total inductance, i.e., the sum of the kinetic and geometric inductances. State-of-the-art superconducting resonators have a quality factor of millions. %
In thin films at low temperature the superfluid contribution to the sheet kinetic inductance is~\cite{diener_homogeneous_2012}:
\begin{eqnarray}
\label{Lk_Drude}
L_K=\frac{m}{2n_se^2d}
\end{eqnarray}
where  $m$ is the effective electron mass,  $d$  is the superconductor thickness, $n_s$ is the superfluid density and  $e$ is the elementary charge. Equation~\eqref{Lk_Drude} highlights a key principle for the readout scheme: a decrease of the superfluid density $n_s$ results in an increase of the kinetic inductance $L_K$, and thus in a change of the resonance frequency. Since the circulating current in a superconductor decreases the superfluid density $n_s$~\cite{gennes_superconductivity_nodate},  the superconducting resonator frequency can be adjusted using a small bias current. The first demonstration of this concept, for one resonator, has been achieved by M. R. Vissers and collaborators in 2015~\cite{vissers_frequency-tunable_2015}. 
 
The current generated by a thermoelectric detector will be sent into a superconducting resonator. In turn, the resonance frequency will vary with the current. For a current $I$ small compared to the critical current $I_*$ the relative frequency shift is 
 \begin{eqnarray}
 \label{sensitivity_iKID}
 \frac{\delta f}{f}=-\frac{\alpha}{2}\left(\frac{I}{I_*}\right)^2,
 \end{eqnarray}
where $\alpha$ is the fraction of the kinetic inductance over the total inductance (kinetic plus geometric) at zero bias current. Detection of a small current then requires a small critical current $I_*$. For this purpose, two parameters can be optimized: the material dependent critical current density $J_*=I_*/S$, and the cross section $S$ of the superconducting wire. Small critical current density can be achieved especially in disordered superconductor materials, such as granular aluminium, for which $J_*\sim3\times10^{10}$~ A$\cdot$m$^{-2}$ \cite{buisson_observation_1994,dupre_tunable_2017}. Using electronic e-beam lithography nanowire width as small as 100~nm can be obtained for superconducting film thickness of the order of 50~nm. Eventually, we estimate that for superconducting granular aluminum %
it is possible to achieve critical current as low as 150~$\mu$A. For a resonance at 1~GHz such critical current leads to a frequency shift of 14~kHz for an injected current of 800~pA.  In principle, it is thus possible to detect the 800~pA generated by the thermoelectric detectors  for typical millimeter detection application.
 
The general problem is to design a resonator that achieves a high quality factor even when connected to detectors with low impedance (less than one Ohm). To achieve this, the dc-current lines must act as, or go through, a low-pass filter to allow the dc-current (DC-kHz) to flow from the detector to the resonator but prevent the resonating ac-current (GHz range) to leak out of the resonator. We have explored different options, including planar superinductances \cite{Masluk2012}, 3-D lumped-element capacitors or inductive-grounded coplanar waveguide configurations. We describe in the following the first two options.

\begin{figure}[t]
\begin{center}
\resizebox{8cm}{!}{\includegraphics{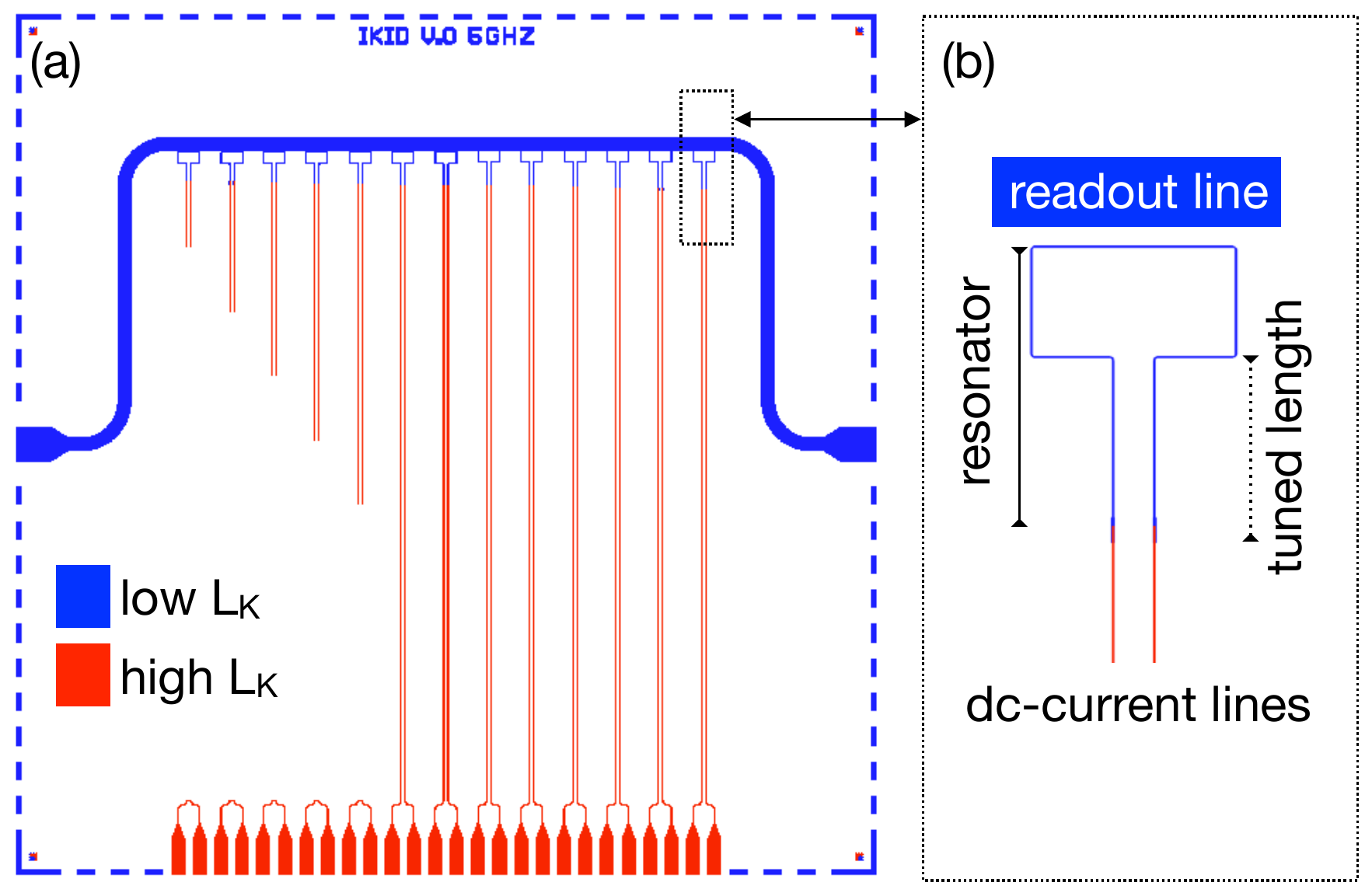}}
\caption{\textbf{Design of the iKID readout scheme with two different materials design} (a) Resonators (in blue) are to connect to the detectors (not shown) via the dc-current lines (in red). The blue horizontal line is a readout line  to monitor the resonance frequencies. (b) Zoom on a resonator.}
\label{ikid_design}
\end{center}
\end{figure}

\subsubsection{Planar superinductances}
We use two different granular aluminium (grAl) compositions: a low kinetic inductance grAl for the resonator part (in blue in Fig.~\ref{ikid_design}) and a high kinetic inductance grAl (in red in Fig.~\ref{ikid_design}) for the two wires connecting the resonator to the detector (superinductances). 
Figure~\ref{ikid_design} shows the design of the readout scheme for current bias kinetic inductance detectors (iKIDs). We  designed thirteen resonators. For testing purposes, only eight were connected to DC-current ports. The  readout line  (blue horizontal line) allows monitoring simultaneously the resonance frequencies. The resonance frequencies are adjusted by varying the total length of the resonators. The electromagnetic simulations foresee resonance frequencies in the 5-7~GHz range and quality factor in the  3-8$\times10^{4}$ range.

\begin{figure}[t]
\begin{center}
\resizebox{8cm}{!}{\includegraphics{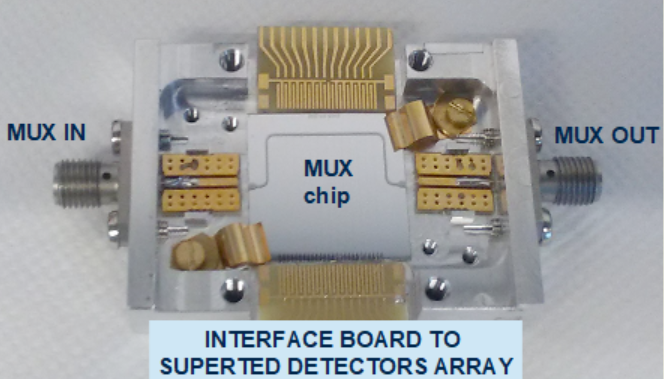}}
\caption{Picture of the iKID readout scheme prototype with two different materials. The matrix of thirteen resonators is in the center. Only the readout line is clearly visible: it is the horizontal line. The bottom PCB-board is connected to the resonators and to the cables.}
\label{ikid_photo}
\end{center}
\end{figure}

An iKID prototype chip was realised and packaged in collaboration with I.~Pop and P.~Paluch from Karlsruhe Institute of Technology (KIT), Germany. Figure~\ref{ikid_photo} shows the prototype inserted in a specially designed sample holder with an ad-hoc PCB-board. The PCB-board (at the bottom) is connected on one side to the superinductances via micro-bonding and on the other side to input DC cables. The prototype was cooled down at 100~mK with cables connected to a dc-current source at room temperature. The measurements demonstrate a sensitivity of the resonance frequency upon the circulating current but most probably due to Joule heating instead of kinetic inductance variation as the quality factors were concomitantly reduced. The Joule heating may be due to the high resistive contacts between the two different materials.

\subsubsection{3-D lumped-element capacitance}
\begin{figure}[t]
\begin{center}
\resizebox{8cm}{!}{\includegraphics{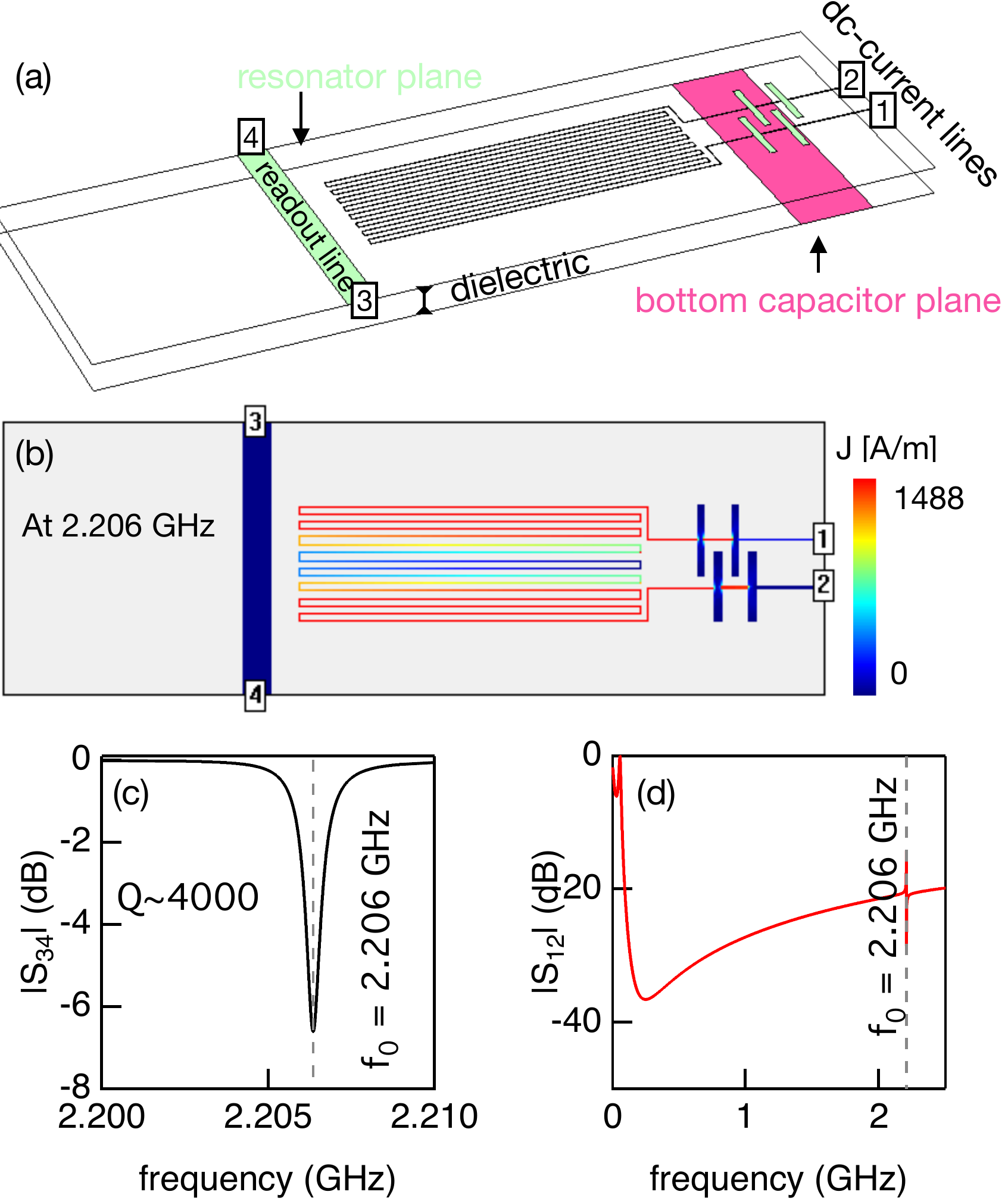}}
\caption{Electromagnetic simulation of a promising configuration of iKID based on 3-D lumped-element capacitance. (a) 3D-model. The dielectric is a 30~nm thick alumina. The metal, green and pink, is aluminum. (b) Current distribution at the resonance. (c) Electrical response of the KID. The quality factor of the resonance is of about $4\cdot10^3$. (d) Electrical response of the DC-lines. The DC-lines are low pass filters.}
\label{ikid_next}
\end{center}
\end{figure}

In this case, both ends of the MUX resonator are grounded (at high frequency) by dielectric parallel-plate capacitors. The size of each capacitor is of about $100\times 500$ $\mu m^2$, smaller than the resonator, assuming a 30~nm thick alumina as dielectric with $\epsilon \approx 10$. In Fig.~\ref{ikid_next} we show the electromagnetic simulation of the model. The resonance exhibits a promising quality factor of the order of about $4\cdot10^3$, paving the way to achieving a multiplexing factor, i.e. the number of detectors that can be read out on a single wire, of hundreds to thousands.

In summary, 
based on the first prototype results, thoughts on the new design and previous experience with disordered superconductors \cite{Levy2019}, we conclude that an iKID should be realised using a single (patterned) layer of granular Aluminium near the insulating transition. This should be capped with a 30~nm thick alumina layer and a second superconducting metal layer, e.g. standard aluminium, to realise the low-pass filter. The quality factor of the resonating readout circuits will allow achieving a multiplexing factor of hundreds to thousands.

\section{Discussion and outlook}\label{sec12}

{In this review, we have presented the fundamentals of a new type of a detector based on thermoelectric and non-reciprocal effects. These arise from the interplay of two phenomena:  The spin-splitting of the density of states of a superconductor via the magnetic proximity effect, and spin polarization in transport through tunnel barriers. 

In the first part of the text we focus on studying the hybrid thin film structure composed of the ferromagnetic insulator EuS adjacent to a superconducting Al. As discussed in Sec.~\ref{sec1}-\ref{sec4}, such  EuS/Al-based devices can be used beyond detectors as non-reciprocal elements for low temperature electronics, 
such as diodes\cite{strambini2022superconducting,diodepatent}. We can also envisage  more elaborated circuits including mixers, reverse current regulators, voltage clamping, and other passive superconducting electronics\cite{braginski2019superconductor}, with the additional control provided by magnetic fields.

Optimizing these devices requires increasing spin polarization using new combinations of materials with higher spin-filter efficiency and reducing their size. An interesting future direction is to use  magnetic two-dimensional van der Waals materials either as sources of the magnetic proximity effect or as spin filtering tunneling barriers.
Interestingly, the non-reciprocity of transport in these devices is a new way to characterize the magnetic proximity effect and spin polarization, an alternative to tunnel spectroscopy. In particular, when the induced spin splitting field cannot be directly resolved from the presence of the BCS peaks in tunneling characteristics, its presence can be determined from the non-reciprocal characteristics of the device.

In this text we have concentrated on the non-reciprocal quasiparticle tunneling characteristics. Besides that, in the presence of either spin-orbit coupling or position dependent exchange field, also the supercurrent becomes non-reciprocal \cite{nadeem2023superconducting}. Such effects form a complementary way of characterizing the magnetic proximity effect.

In the second part, we concentrate on the properties of the SFTED. Specifically, in Sec.~\ref{sec5}, we review the microcalorimetry theory of the   SFTED detectors and determine their  noise-equivalent power and energy resolution as a function of the system parameters. We also discuss in this section possible designs of  the SFTED for the X-ray detection.
Section \ref{sec6} briefly discusses the THz range accessible for SFTED, and present preliminary measurements  of a   SFTED junction. 
Finally, in Sec.~\ref{sec7} we discuss  possibilities of read-out and multiplexing. 
}

The detector work also highlights challenges in fabricating the SFTED device. It would be desirable to have a detector controllable only with a rather small magnetic field, as the coil used to create it may complicate coupling of the radiation into the detector. Ideally no field would be required at all and one would rely on the intrinsic magnetization of the ferromagnetic insulators. However, for example for EuS the Curie temperature is rather low (about 14-17 K, depending on doping), and it needs to be magnetized after the sample has been refrigerated to the base temperature. Besides coils, this could perhaps be done using permanent magnets aligned with the sample.

To minimize the required magnetic field, and thereby to ensure optimal magnetic proximity effect, we have prepared the EuS/Al systems using in-situ shadow mass patterning without breaking the vacuum. Such a process results to relatively large junctions, with lateral dimensions of the order of a few hundred $\mu$m$^2$. Optimal THz detectors would constitute planar antennas with small-size absorbers and small-size junctions (much smaller than the wavelength), and further size reduction is desired. In the X-ray regime the energy resolution is mostly determined by the relaxation time of the radiation absorber, which is inevitably significant, so the mere size of the junction is not a problem. However, in that regime the detection requires fairly transparent junctions, i.e., with a low normal-state tunneling resistance $R_T$ (see Fig.~\ref{fig:xray_dEvsRandT}). This requirement is not straightforward to realize in large junctions without the presence of pinholes and the resulting large number of subgap states, quantified with the parameter $\Gamma$. Ideal detector would also require rather low $\Gamma \lesssim 10^{-2} \Delta$ as indicated in Fig.~\ref{fig:xray_dEvsParams}. Moreover, %
an improved energy resolution for the detection of keV X-rays could possibly be obtained by suspending the absorber to prevent the premature escape of athermal and thermal phonons\cite{nahum1995,kozorezov2013}. Such design has been widely adopted for transition edge sensors\cite{Ullom2015}.
On the other hand, the junctions themselves should not reside in the suspended region, as maximizing the temperature difference would require thermalizing the normal part of the junction as well as possible.

We are confident that these technical challenges have a solution. For example, it may be better to pay the price of a sub-optimal FI/S interface to get a nanostructured detector. Once such detectors are in place, SFTED-based detectors can rival existing superconducting detectors such as transition edge sensors and kinetic inductance detectors. Subsequantly, the next step would be to construct a large number of detectors and replace the current read-out with a multiplexing technique, such as the nanowire based multiplexing discussed in Sec.~\ref{sec7}. One of the issues to be handled in such an SFTED detector array would be the dipolar coupling between the magnetizations. However, as the performance of the detector does not rely on the possibility of (locally) switching magnetization from one state to another, the whole array could be controlled with a single magnetic field. %

Besides detectors and nonreciprocal electronics, superconductor-ferromagnet multilayers enable other novel phenomena. The spin-split superconductors can be used to create absolute spin valves\cite{de_simoni_toward_2018}, thermal logic circuits\cite{paolucci2018phase}, coherent caloritronic systems\cite{paolucci2017phase,giazotto2014proposal,giazotto2015very}, devices mixing nonequilibrium spin and equilibrium supercurrent transport \cite{rezaei2020phase} and thermal quantum devices\cite{marchegiani2016self,giazotto2015ferromagnetic,giazotto2020very}. Moreover, a rich dynamical response of the hybrid systems can be envisaged: The presence of superconductors influence the magnetic resonance by mediating magnetic interactions (as a consequence of the magnetic proximity effect) \cite{degennes1966-cfs,ojajarvi2022dynamics} and via the modified spin battery effect. Moreover, they may give access to coupling different collective modes, those inherent to magnetism with those relevant for superconductivity \cite{lu2022coupling}. 

We wish  to finally mention that a very large thermoelectric response was  recently observed in fully superconducting tunnel junctions composed of different superconductors\cite{germanese2022bipolar,germanese2023phase}. These systems, although being perfectly reciprocal, exhibit a \emph{spontaneous} breaking of the electron-hole symmetry in the presence of a large thermal bias\cite{marchegiani2020nonlinear,marchegiani2020phase}, thereby providing sizable thermoelectric figures of merit that could be exploited for the implementation of novel-concept nanoscale thermoelectric machines\cite{germanese2022bipolar,germanese2023phase,marchegiani2020superconducting} and sensitive passive radiation detectors\cite{paolucci2023highly}.

\section*{Acknowledgements}
{This work was mainly funded by  EU's Horizon 2020 Research and Innovation Program under Grant Agreement
No.~800923 (SuperTED). F.S.B and C.R also acknowledge funding from the Spanish MICINN-AEI (Projects No.~PID2020-114252GB-I00 (SPIRIT) and TED2021-130292B-C42), and  the Basque Government (grant IT-1591-22),} M.S. and E.S. acknowledge partial funding from the European Union’s Horizon 2020 research and innovation programme under the Marie Skłodowska Curie Action IF Grant No. 101022473 (SuperCONtacts). C. I. L. A. acknowledge partial funding from CNPq, CAPES and Fapemig.
F.G. acknowledges  the EU’s Horizon
2020 research and innovation program under Grant Agreement No. 964398 (SUPERGATE) for partial financial support.

The data of the figures in this work can be obtained at \href{https://doi.org/10.5281/zenodo.7798143}{https://doi.org/10.5281/zenodo.7798143}.

\section*{References}

\bibliographystyle{iopart-num}
\bibliography{refs}

\end{document}